\documentclass[
    traditabstract,
    longauth,
]{aa}
\usepackage[varg]{txfonts}
\usepackage{natbib}
\bibpunct{(}{)}{;}{a}{}{,}
\usepackage{graphicx}
\usepackage{amsmath}
\usepackage{amssymb}
\usepackage{ragged2e}
\usepackage[
    separate-uncertainty=true,
    multi-part-units=single,
    range-phrase=\mbox{--},
    range-units=single
]{siunitx}
\ifdefined\unit\else
  \ifdefined\NewCommandCopy
    \NewCommandCopy\unit\si
  \else
    \NewDocumentCommand\unit{O{}m}{\si[#1]{#2}}
  \fi
\fi
\usepackage{hyperref}
\usepackage{enumitem}
\usepackage{dblfloatfix}
\usepackage{xfrac}

\usepackage{lscape}
\usepackage{multirow}
\usepackage{array}
\newcommand{\PreserveBackslash}[1]{\let\temp=\\#1\let\\=\temp}
\newcolumntype{C}[1]{>{\PreserveBackslash\centering}p{#1}}

\DeclareSIUnit{\au}{AU}
\DeclareSIUnit{\dex}{dex}
\DeclareSIUnit{\hjulianday}{HJD}
\DeclareSIUnit{\Jansky}{Jy}
\DeclareSIUnit{\julianday}{BJD}
\DeclareSIUnit{\mas}{mas}
\DeclareSIUnit{\mag}{mag}
\DeclareSIUnit{\parsec}{pc}
\DeclareSIUnit{\permille}{\text{\textperthousand}}
\DeclareSIUnit{\rad}{rad}
\DeclareSIUnit{\solarmass}{M\textsubscript{\(\odot\)}}
\DeclareSIUnit{\solarradius}{R\textsubscript{\(\odot\)}}
\DeclareSIUnit{\solarluminosity}{L\textsubscript{\(\odot\)}}
\DeclareSIUnit{\earthmass}{M\textsubscript{\(\oplus\)}}
\DeclareSIUnit{\earthflux}{\text{$\Phi$}\textsubscript{\(\oplus\)}}
\DeclareSIUnit{\year}{yr}
\DeclareMathOperator{\avg}{avg} 

\bibpunct{(}{)}{;}{a}{}{,}

\defcitealias{gaiaDR3}{GDR3}
\defcitealias{SuarezMascareno2024}{SM24}
\defcitealias{SuarezMascareno2018}{SM18}
\defcitealias{Affer2016}{A16}
\defcitealias{Dodson-Robinson2022}{DR22}
\defcitealias{Henry1996}{H96}
\defcitealias{Hara2022}{H22}
\defcitealias{GomesDaSilva2011}{GDS11}
\defcitealias{Giles2017}{G17}

\makeatletter
\DeclareRobustCommand
  \shortvdots{\vbox{\baselineskip4\p@ \lineskiplimit\z@
    \hbox{.}\hbox{.}\hbox{.}}}
\makeatother 
\title{HADES RV Programme with HARPS-N at TNG}
\subtitle{XVI. A super-Earth in the habitable zone of the GJ~3998 multi-planet system\thanks{Based on observations made with the Italian \textit{Telescopio Nazionale Galileo} (TNG) operated by the \textit{Fundación Galileo Galilei} (FGG) of the \textit{Istituto Nazionale di Astrofisica} (INAF) at the \textit{Observatorio del Roque de los Muchachos} (La Palma, Canary Islands, Spain).}}

\authorrunning{Stefanov et al.}
\titlerunning{A super-Earth in the habitable zone of the GJ~3998 multi-planet system}
\author{
A. K. Stefanov\inst{1,2,*},
A. Su\'arez Mascare\~no\inst{1,2},
J. I. Gonz\'alez Hern\'andez\inst{1,2},
N. Nari\inst{1,2,3},
R. Rebolo\inst{1,2,4},
L. Affer\inst{5},
G. Micela\inst{5},
I. Ribas\inst{6,7},
A. Sozzetti\inst{8},
M. Perger\inst{6,7},
M. Pinamonti\inst{8},
M. Damasso\inst{8},
J. Maldonado\inst{5},
E. Gonz\'alez \'Alvarez\inst{9},
G. Scandariato\inst{10}
}

\institute{
\inst{1}Instituto de Astrof\'isica de Canarias, 38205 La Laguna, Spain\\
\inst{2}Departamento de Astrof\'isica, Universidad de La Laguna, 38206 La Laguna, Spain\\
\inst{3}Light Bridges S. L., 35004 Las Palmas de Gran Canaria, Spain\\
\inst{4}Consejo Superior de Investigaciones Cient\'ificas (CSIC), 28006 Madrid, Spain\\
\inst{5}INAF - Osservatorio Astronomico di Palermo, Piazza del Parlamento 1, 90134 Palermo, Italy\\
\inst{6}Institut de Ci\`encies de l’Espai (ICE, CSIC), Campus UAB, Carrer de Can Magrans s/n, 08193 Bellaterra, Spain\\
\inst{7}Institut d’Estudis Espacials de Catalunya (IEEC), c/ Gran Capit\`a 2-4, 08034 Barcelona, Spain\\
\inst{8}INAF - Osservatorio Astrofisico di Torino, Via Osservatorio 20, 10025 Pino Torinese, Italy\\
\inst{9}Departamento de F\'isica de la Tierra y Astrof\'isica \& IPARCOS-UCM (Instituto de F\'isica de Part\'iculas y del Cosmos de la UCM), Facultad de Ciencias F\'isicas, Universidad Complutense de Madrid, 28040 Madrid, Spain\\
\inst{10}INAF - Osservatorio Astrofisico di Catania, Via S. Sofia 78, 95123 Catania, Italy\\
\inst{*}\email{atanas.stefanov@iac.es}
}

\date{Received 16 October 2024 / Accepted 30 January 2025}

\abstract{The low masses of M dwarfs create attractive opportunities for exoplanet radial-velocity (RV) detections. These stars, however, exhibit strong stellar activity that may attenuate or mimic planetary signals. We present a velocimetric analysis on one such M dwarf, GJ~3998 \mbox{($d=18.2$\,pc)}, with two published short-period super-Earths: GJ~3998~b and GJ~3998~c. We use additional data from the HARPS-N spectrograph to confirm these two planets and to look for more. We carry out joint modelling of: (i) RV planetary signals, (ii) stellar rotation in RV and activity indicators through Gaussian processes, (iii) long-term trends in RV and activity indicators. We constrain the rotational period of GJ~3998 to \mbox{$P_\text{rot}=30.2\pm 0.3\,$d} and discover long-term sinusoidal imprints in RV and FWHM of period \mbox{$P_\text{cyc}=316^{+14}_{-8}$\,d}. We confirm GJ~3998~b and GJ~3998~c, and detect a third planet: GJ~3998~d, whose signal had been previously attributed to stellar activity. GJ~3998~d has an orbital period of $41.78\pm 0.05$\,d, a minimum mass of \mbox{$6.07^{+1.00}_{-0.96}\,$M\textsubscript{\(\oplus\)}} and a mean insolation flux of \mbox{$1.2^{+0.3}_{-0.2}\,$\text{$\Phi$}\textsubscript{\(\oplus\)}}. This makes it one of the few known planets receiving Earth-like insolation flux.}


\begin{document}

\maketitle

\section{Introduction}
Exoplanetary astronomy promises to be more fruitful than ever, with the number of confirmed planets recently having exceeded 5800 (\citealp{nasaexoplanettable}; accessed Jan 2025). The rate of detection is still on the rise owing to many transit-photometry surveys, most notably Kepler \citep{kepler} and TESS \citep{tess}. Although now behind in number of detections, the method of radial velocities (RVs), which once initiated the search for planets with \citet{Mayor1995}, continues to deliver. The RV method remains a favoured technique because: (i) it is less selective in terms of system geometry; (ii) it is convenient for follow-up confirmation of planet candidates, which is the case for many Kepler and TESS targets; and (iii) it allows for the masses of exoplanets to be constrained, which provides information on their interiors, as long as their radii are determined. Moreover, deep RV searches are very desirable for the nearest stars, since only a small fraction of planets are generally expected to transit.

The RV method measures the gravitational influence on a star from one or more other bodies. This suggests that M dwarfs -- the least massive and most prevalent main-sequence stars in the Galaxy -- are the theoretical favourites for this technique. Occurrence-rate surveys have indicated that M dwarfs host at least one planet on average \citep{Dressing2015,Tuomi2019,Hsu2020,Sabotta2021,Perger2023}. In practice, however,
M dwarfs stubbornly mask potential planetary RV signals through physical processes inherent to late-type stars, most notably stellar activity on both short and long timescales (e.g. \citealp{Borgniet2015} and \citealp{Dumusque2011a}, respectively). In this work, we examine the RV behaviour of one such M dwarf, GJ~3998 \mbox{($\alpha=17\textsuperscript{h}16\textsuperscript{m}00.6\textsuperscript{s}$}, 
\mbox{$\delta=11\degr03'27.6''$}; \citealp{gaiaDR3}, hereafter \citetalias{gaiaDR3}). The distance to GJ~3998 has been constrained to \SI{18.2}{\parsec} \citepalias{gaiaDR3}, and its stellar parameters have been reasonably well established (Table~\ref{tab:stellar_parameters}). Most recently, GJ~3998 has been studied by the HADES RV programme that was carried out at the HARPS-N spectrograph at the \SI{3.58}{\metre} \textit{Telescopio Nazionale Galileo} (TNG; \citealp{harpsn}). \citet{Affer2016}, hereafter \citetalias{Affer2016}, modelled the stellar activity that GJ~3998 exhibited, and found evidence of two planetary companions around it:
GJ~3998~b, with a period near \SI{2.65}{\day}, an eccentricity of zero, and a minimum mass of \mbox{$2.47\pm 0.27\,\unit\earthmass$}; and
GJ~3998~c, with a period near \SI{13.74}{\day}, an eccentricity of $0.049^{+0.052}_{-0.034}$, and a minimum mass of $6.26^{+0.79}_{-0.76}\,\unit\earthmass$. The rotational period of GJ~3998 was also determined by \citetalias{Affer2016} to be $P_\text{rot}=31.8^{+0.6}_{-0.5}\,\unit\day$; soon after, a photometric confirmation followed (\citealp{Giacobbe2020}; \SI{32.9925}{\day}).

Since the discovery of GJ~3998~b and GJ~3998~c, the exoplanetary detection field has been enjoying the fruits of novel advancements. These include the use of multi-dimensional Gaussian-process regression \citep{Rajpaul2015, spleaf1, spleaf2} and more sophisticated RV determination techniques (e.g. \citealp{Artigau2022}). Such advancements can lead to the discovery of previously uncaught planetary signals. This circumstance and the proximity of this system constitute our main motivation for a more modern RV analysis of GJ~3998 that includes contemporary data. This work continues as follows. Section~\ref{sec:observations} describes the available observational data relevant to our study. Section~\ref{sec:modelling} narrates the generalised modelling approach that arose over the course of our investigation. Section~\ref{sec:analysis} contains our analysis of available velocimetry. We discuss our results in Sect.~\ref{sec:discussion}, and we summarise them in Sect.~\ref{sec:conclusions}.

\begin{table}
\centering
\caption{Stellar parameters of GJ~3998.}
\label{tab:stellar_parameters}
\begin{tabular}{lclc}
\hline\hline
\multicolumn{1}{c}{Parameter} &
\multicolumn{1}{c}{Unit} &
\multicolumn{1}{c}{Value} &
\multicolumn{1}{c}{Ref.} \\ \hline
$\alpha$ (J2000) &
-- &
17\textsuperscript{h}16\textsuperscript{m}00.6\textsuperscript{s} &
1 \\
$\delta$ (J2000) &
-- &
+11\degr03'27.6'' &
1 \\
$\mu_\alpha\cos\delta$ &
\unit{\mas\per\year} &
$-137.435\pm 0.027$ &
1 \\
$\mu_\delta$ &
\unit{\mas\per\year} &
$-347.456\pm 0.023$ &
1 \\
$\varpi$ &
\unit{\mas} &
$55.0169\pm 0.0287$ &
1 \\
$m_\text{J}$ &
\unit{\mag} &
$7.634\pm 0.021$ &
2 \\
$m_\text{H}$ &
\unit{\mag} &
$7.020\pm 0.029$ &
2 \\
$m_\text{K}$ &
\unit{\mag} &
$6.816\pm 0.016$ &
2 \\
Type &
-- &
M1.0 &
3 \\
$T_\text{eff}$ &
\unit{\kelvin} &
\num{3726(68)} &
3 \\
$M_\star$ &
\unit{\solarmass} &
$0.52\pm 0.05$ &
3 \\
$R_\star$ &
\unit{\solarradius} &
$0.50\pm 0.05$&
3 \\
$\log_{10}L_\star$&
\unit\solarluminosity &
$-1.358\pm 0.086$ &
3 \\
$\log_{10}g$&
\unit{\centi\metre\per\second\squared} &
$4.75\pm 0.04$ &
3 \\
Age &
\unit{\giga\year} &
$8.38\pm 4.06$ &
3 \\
\text{[Fe/H]} &
\unit{\dex} &
$-0.13\pm 0.09$ &
3 \\
\hline
\end{tabular}
\tablebib{
(1) \citet{gaiaDR3};
(2) \citet{2mass};
(3) \citet{Maldonado2020}
}
\end{table} 
\section{Observations}\label{sec:observations}
\subsection{HARPS-N velocimetry}\label{sec:harpsn_velocimetry}
HARPS-N is a fibre-fed high-resolution échelle spectrograph installed at the \SI{3.58}{\metre} TNG at the Roque de los Muchachos Observatory, Spain \citep{harpsn}. It has a resolving power of $R\sim\num{115 000}$ over a \SIrange{380}{690}{\nano\metre} passband. \mbox{HARPS-N} is contained in a vacuum vessel to limit spectral drifts induced by temperature- and air pressure variations. This ensures its stability with the course of time. HARPS-N comes with its own pipeline that provides wavelength-calibrated spectra, as well as higher-level data products, including: reduced two- and one-dimensional wavelength-calibrated spectra, cross-correlation functions (CCFs), CCF bisector profiles and CCF-derived RVs. Like for other HADES targets, GJ~3998 was not observed with its own simultaneous thorium-argon (ThAr) lamp calibration. Because of this, our data is not corrected for the instrumental drift between the star fibre and the reference-calibration fibre. This drift is expected to have a mean variance of \SI{1.00}{\metre\per\second} \citep{Perger2017}. Thus, we anticipate a slight inflation in the RV model jitter.

\subsubsection{Radial velocities}\label{sec:obs_rvs}

HARPS-N CCF RVs form a good baseline for analysis, but RV determination can be rendered even more precise through other techniques on raw spectra. The line-by-line method (LBL method; \citealp{Artigau2022}) offers a new mode of RV extraction that uses the full spectral information to measure velocities and that handles outliers in a statistically consistent manner. In addition, the LBL method provides with many complementary metrics that give insights of stellar activity, such as differential line widths, or even RVs measured from specific wavelength ranges.

We reduced 206 available raw HARPS-N spectra through the YABI portal hosted by the Centro Italiano Archivi Astronomici. YABI \citep{yabi1,yabi2} implements the classical HARPS-N pipeline and computes CCFs for a given template mask. We used an M2-type mask to obtain CCF RVs. Then, we fed all raw spectra to the LBL method with automatic telluric cleaning and an effective-temperature estimate informed by Table~\ref{tab:stellar_parameters}.\footnote{We used LBL v0.65.001 with the abbreviated commit hash \texttt{2fabec6}.} We moved forwards with LBL RVs and several curated activity indicators (Sect.~\ref{sec:activity_indicators}). One point near \SI{2457124.7}{\julianday} was rejected by LBL because it did not meet its \mbox{$\text{S/N}>8$} requirement. We discarded another point near \SI{2456694.7}{\julianday} on account of being a strong outlier in three activity indicators. This returned a dataset of 204 points that spans from May 2013 to Sep 2023 and has a median temporal spacing of \SI{1.0}{\day}. This final dataset, which we carried forwards in subsequent modelling and analysis, is 50\% greater in size than \citetalias{Affer2016} ($N=136$). It has an LBL RV root mean square (rms) of \SI{4.28}{\metre\per\second}, which is noticeably lower in comparison to CCF RVs (\SI{4.98}{\metre\per\second}). Moreover, the RV precision appears to have improved significantly -- we observe a median LBL RV uncertainty of \SI{1.02}{\metre\per\second}, compared to \SI{1.71}{\metre\per\second} for CCF RVs.

\subsubsection{Activity indicators}\label{sec:activity_indicators}
We took to considering four metrics that quantify stellar activity in their own way: CCF FWHM, the Mount Wilson S-index, the H$\alpha$ line and the Na~I~D lines. We collectively refer to those metrics as `activity indicators'.

Active regions perturb the flux- and velocity fields of the stellar disc, thereby changing the shape of observed lines. Consequently, the CCF also changes in shape. Such perturbations are quantified by tracking the evolution of CCF width, depth, and symmetry, and their strength can be related to the coverage of the active regions, their contrast and the stellar $v\sin i$. In this work, we use the CCF full width at half maximum (FWHM) as a primary activity indicator. The CCF FWHMs are automatically provided by the HARPS-N DRS; we took them and then colour-corrected them order by order following \citet{SuarezMascareno2023}. Our considered CCF FWHM time series are described by an rms of \SI{5.77}{\metre\per\second} and a median uncertainty of \SI{3.42}{\metre\per\second}.

The emission intensity of the cores of Ca II H\&K lines is a reliable proxy for the strength of the stellar magnetic field, and thereby of the stellar-rotation period, $P_\text{rot}$, for late-type stars \citep{Noyes1984,SuarezMascareno2015}. We used the Mount Wilson \mbox{S-index},
\begin{equation}
    S_\text{MW} = 1.111\times\frac{\Tilde{N}_\text{H}+\Tilde{N}_\text{K}}{\Tilde{N}_\text{R}+\Tilde{N}_\text{V}}+0.0153,
\end{equation}
where $\Tilde{N}_\text{H}$ and $\Tilde{N}_\text{K}$ are triangular passbands centred at \SI{3968.470}{\angstrom} and \SI{3933.664}{\angstrom}, with a common FWHM of \SI{1.09}{\angstrom}; while $\Tilde{N}_\text{R}$ and $\Tilde{N}_\text{V}$ are rectangular passbands centred at \SI{4001.07}{\angstrom} and \SI{3901.07}{\angstrom}, with a common width of \SI{20}{\angstrom} \citep{Vaughan1978}. Our considered measurements of $S_\text{MW}$ are described by an rms of \num{0.133} and a median uncertainty of \num{0.077}.

The H$\alpha$ line and the Na~I~D lines are also good activity indicators for low-activity, low-mass stars. We computed H$\alpha$ and Na~I~D fluxes by the definitions of \citet{GomesDaSilva2011} and \citet{Diaz2007}, respectively. In our considered measurements, the H$\alpha$ indicator is characterised by a rms of \num{2.44e-2} and a median uncertainty of \num{0.27e-2}. This is potentially indicative of strong activity relative to the instrumental precision. The Na~I indicator has a rms of \num{4.71e-3} and a median uncertainty of \num{1.53e-3}. Uncertainties of all activity indicators except CCF FWHM come from photon-noise propagation in their algebraic expressions.

\subsection{HARPS velocimetry}
HARPS is a high-resolution échelle spectrograph located at the \SI{3.6}{\metre} ESO telescope, La Silla Observatory, Chile \citep{harps}. The public HARPS RV database by \citet{harps_trifon} contains 6 RV measurements of GJ~3998. Acquired measurements span from April 2008 to July 2014, with a median temporal difference of \SI{1.0}{\day}. Compared to the HARPS-N collection, HARPS data are too small in number and too isolated in epoch to justify an inclusion in our analysis.

\subsection{ASAS-SN photometry}

ASAS-SN is an automated photometric programme that looks out for supernovae and other transient events \citep{asassn}. It is comprised of 24 ground-based \SI{14}{\centi\metre} telescopes that are grouped in fours at six sites, all part of the Las Cumbres Observatory Network \citep{Brown2013}. Sites are located at: Haleakalā Observatory, HI, USA; two at Cerro Tololo Observatory, Coquimbo, Chile; South African Astronomical Observatory, Sutherland, South Africa; McDonald Observatory, TX, USA; Ali Observatory, Tibet, China. This heterogeneous distribution of the telescope network and the large number of unit telescopes allow ASAS-SN to survey the entire visible sky every night, with an uptime that is less prone to weather conditions. We present a short self-contained photometry analysis in Appendix~\ref{sec:photometry_analysis}. Its key takeaway is that we find no signals near periods of discussed planetary companions.

\section{Modelling and inference}\label{sec:modelling}
We now describe our mode of operation that is generalised for $N$ data sources (e.g. HARPS-N, HARPS etc.) and $M$ physical quantities (e.g. RV, FWHM etc.). We denote measurements $y_{i,j}$, their associated uncertainties $\Delta y_{i,j}$, and the times of measurement $t_{i,j}$, where $i\in[0..N-1]$ and $j\in[0..M-1]$. We imposed two requirements: 
\mbox{(i) $\forall i,j:\avg y_{i,j}=0$},
\mbox{(ii) $\max{t_{i,j}}=0$}.
The first condition establishes common ground for the introduction of instrument offsets (Sect.~\ref{sec:modelling_offsets_jitters}).
The second condition minimises uncertainty on published ephemerides, and avoids complex degeneracies between parameters of periodic long-term corrections (Sect.~\ref{sec:modelling_correctors}).
Some additional preliminary steps need to be made in order to clean data of instrumental or physical trends. Then, the stellar activity was modelled together with potential planetary signals. The following subsections narrate the steps of pre-processing, stellar-activity modelling and the optional planetary modelling. Our models assume that stellar activity influences all considered physical quantities, while potential planets act on RV only (conventionally assigned to $j=0$).

\subsection{Source offsets and jitters}
In cases where spectra were acquired from different instruments, we accounted for non-common zero-points through the substitution
\label{sec:modelling_offsets_jitters}
\begin{equation}
    y_{i,j}\to
    y_{i,j}-O_{i,j}\qquad
    \begin{array}{l}
        \forall\, i\neq 0, \\
        \forall j,
    \end{array}
    \label{eq:source_offset}
\end{equation}
where $O_{i,j}$ is the offset of source $i$, in the quantity $j$. \mbox{HARPS-N} is the zero-point and the only source in this analysis.

Beyond instrumental precision, measurements come with additional uncorrelated noise that we are in principle unable to model. We accounted for this by adding jitter terms in quadrature to measurement uncertainties; that is, through the substitution
\begin{equation}
    \Delta y_{i,j}\to
    \sqrt{\left(\Delta y_{i,j}\right)^2+J^2_{i,j}}\qquad
    \begin{array}{l}
        \forall\, i, \\
        \forall j,
    \end{array}
    \label{eq:source_jitter}
\end{equation}
where $J_{i,j}$ is the instrumental jitter of source $i$ in quantity $j$.

\subsection{Long-term correction}\label{sec:modelling_correctors}
Radial velocity extraction pipelines sometimes provide with data that follow strong trends. Said trends may be physically motivated (e.g. secular acceleration or long-term stellar cycles), but they may also be systematics uncaught by reduction pipelines (e.g. instrumental drifts in RV). We applied corrections through the substitution
\begin{equation}
    y_{i,j}\to
    y_{i,j}-f_j(t_{i,j})\qquad
    \begin{array}{l}
        \forall\, i, \\
        \forall j,
    \end{array}
    \label{eq:longterm_correction}
\end{equation}
where $f_j$ is the long-term function (LTF) for quantity $j$. LTFs may or may not share parameters across quantities.

\subsection{Planetary signatures}\label{sec:modelling_planets}
Searches for planetary signatures require comparisons between models with and without planets. For models with planets, we subtracted each planetary signal in RV through the substitution
\begin{equation}
    y_{i,0}\to
    y_{i,0}-k_\text{rv}
    \left\{
    \cos\left[\omega+\nu(t_{i,0})\right]+e\cos\omega
    \right\}
    \qquad
    \forall i,
    \label{eq:planetary_subtraction}
\end{equation}
where $k_\text{rv}$ is the RV semi-amplitude of the signal, $\omega$ is the argument of periastron, $\nu$ is the true anomaly function, and $e$ is the planetary eccentricity. Planetary orbits are characterised by five parameters: $k_\text{rv}$, $e$, $\omega$, the period, $P$, and the phase, $\varphi$. We defined $\varphi$ such that $\max(t_{i,j})$ has a normalised phase $\varphi$ relative to the latest inferior conjunction, as was defined in \citet{Kane2009} under the name `midpoint of primary transit'. Inferior-conjunction ephemerides are therefore located at
\begin{equation}
    \varepsilon=(n-\varphi)P\qquad \forall n\in\mathbb{Z}.
\end{equation}

To solve \eqref{eq:planetary_subtraction}, we utilised the \textsc{EKEPL1} procedure by \citet{Odell1986}, which was reported to achieve precision better than \SI{e-12}{\rad} after five iterations.
Nevertheless, it is sometimes favourable to approximate \eqref{eq:planetary_subtraction} to
\begin{equation}
    y_{i,0}\to
    y_{i,0}+k_\text{rv}
    \sin\left[\
    \frac{2\pi(t_{i,0}+P\varphi)}{P}
    \right]
    \qquad
    \forall i,
    \label{eq:plasine_subtraction}
\end{equation}
which is much faster to compute, and which introduces only three parameters: $k_\text{rv}$, $P$ and $\varphi$. We hereafter refer to \eqref{eq:plasine_subtraction} as the planetary-sine approximation, or the circular-orbit approximation. Here, $\varphi$ is defined as in \eqref{eq:planetary_subtraction}. We note that this preservation of meaning required a change in sign relative to \eqref{eq:planetary_subtraction}.

\subsection{Stellar-activity modelling}\label{sec:modelling_stellar_activity}
Gaussian processes (GPs) lay the foundation of our stellar-activity modelling. Formally, a GP is defined as `a collection of random variables, any finite number of which have a joint Gaussian distribution' \citep{Rasmussen2006}. As GPs offer non-parametric approaches to fit data, observations themselves determine the best functional form of the model, no matter how intricate \citep{Haywood2015}. This makes GP-based frameworks convenient tools for analysing stellar activity in RV, because they require no prior knowledge of associated active regions. GPs have already demonstrated to reap notable successes in planet detections (e.g. \citealp{Haywood2014}). However, the main strength of GPs is also their greatest weakness -- at times, they are flexible enough to overfit data, and in turn to suppress signals of physical nature \citep{SuarezMascareno2023}.

Stellar activity can be modelled as a stochastic process that is characterised by a periodic component (stellar rotation), as well as an exponential component (timescale of coherence of the periodic component). For this purpose, we employed the squared-exponential periodic (SEP) kernel,\footnote{Other works refer to this kernel as the quasi-periodic (QP) kernel.} which is standard treatment of our scientific problem (e.g. \citealp{Haywood2014,Angus2018}). The SEP kernel has the form
\begin{equation}
    k(\Delta t) = \kappa^2\exp\left[
    -\frac{\Delta t^2}{2\tau^2}
    -\frac{\sin^2\left(\pi\Delta t/P\right)}{2\eta^2}
    \right],
    \label{eq:sep_kernel}
\end{equation}
where $\kappa^2$ is the maximum amplitude of the kernel, $\tau$ is the evolutionary timescale, $P$ is the kernel period that we associate with the stellar-rotation period $P_\text{rot}$, and $\eta$ describes the kernel feature complexity. The hyperparameter $\eta$ is parametrised differently in the literature, and its variations have different names; for example, lengthscale or harmonic complexity. We hereafter refer to $\eta$ as the `sinescale' by virtue of its algebraic analogy to the timescale $\tau$. The \textsc{s+leaf} library \citep{spleaf1,spleaf2} provides with many approximations of the SEP kernel, two of which we work with: the Matérn 3/2 exponential periodic (MEP) kernel\footnote{Not to be confused with the Matérn 3/2 kernel.} and the exponential-sine periodic (ESP) kernel.

GP kernels are used in different regimes. One of the commonly used ones, the multi-dimensional regime, tries to fit one GP kernel on all physical quantities at the same time. This is done by solving the system of equations
\begin{equation}
    \left|\begin{array}{l}
    y_{i,j}=A_j\,G(t_{i,j})+B_j\,(\mathrm{d}G/\mathrm{d}t)_{t=t_{i,j}}\\[0.2ex]
    \shortvdots
    \end{array}\right.\qquad
    \begin{array}{l}
        \forall\, i, \\
        \forall j,
    \end{array}
    \label{eq:multidimensional_system}
\end{equation}
where $A_j$ and $B_j$ are quantity-specific fit parameters, and $G(t_{i,j})$ is the kernel estimation of the stellar activity at a given time. The multi-dimensional regime follows the $FF'$ formalism that was described in \citet{Aigrain2012} and extended in \cite{Rajpaul2015}. This regime is a common approach for the disentanglement of stellar activity in the velocimetry of late-type dwarfs, such as K2-233 \citep{Barragan2023} and GJ~9827 \citep{Passegger2024}, and it is useful to break degeneracies in cases where there is a significant correlation between physical quantities \citep{SuarezMascareno2020}. \citet{Rajpaul2015} discussed that the $FF'$ formalism expects only some activity indicators to be `gradiented'; that is, to have a non-zero $B_j$ term in \eqref{eq:multidimensional_system} for $j\neq 0$. We used the \textsc{s+leaf} MultiSeriesKernel procedure to realise the multi-dimensional regime. We set \mbox{$B_j=0$} for each non-gradiented quantity and excluded it from inference.

\subsection{Parameter priors}\label{sec:parameter_priors}
We use a set of default parameter priors unless stated explicitly. This establishes a common baseline for comparison, and lightens the description of the many models we present. 

Our default LTF parameter priors are informed by the solutions of the Levenberg-Marquadt algorithm (LMA; \citealp{Levenberg1944,Marquardt1963}). We ran an instance of the LMA that returns a mean $\mu_\text{lm}$ and an uncertainty $\sigma_\text{lm}$ for each LTF parameter. For these parameters, we then used the priors $\mathcal{N}\left(\mu_\text{lm},200\sigma_\text{lm}\right)$. The large scale factor of $\sigma_\text{lm}$ accounts for the minute errors that the LMA tends to return. Our default offset priors are based on the common standard deviation of all $y_{i,j}$, which we denote $\sigma_j$. The default prior on offsets is \mbox{$O_{i,j}=\mathcal{N}(0,5\sigma_j)$}. This gives wide enough priors in most cases. For jitters, we used \mbox{$J_{i,j}=\mathcal{U}_\text{log}(10^{-3},10^{3})$}, which exploits that jitters are positive by definition.

The default priors on stellar-activity- and planetary parameters were chosen to be non-restrictive. Planets have default RV semi-amplitude priors of
\mbox{$\mathcal{U}_\text{log}(10^{-3},10^3)\,\unit{\metre\per\second}$}
and period priors of
\mbox{$\mathcal{U}_\text{log}(1,10^{3})\,\unit\day$}.
The remaining parameters, $\varphi$, $e$, and $\omega$, have uniform priors that are defined by their domains. For SEP kernel hyperparameters, we used timescale priors of
\mbox{$\mathcal{U}_\text{log}(40,10^{4})\,\unit\day$},
period priors of
\mbox{$\mathcal{U}_\text{log}(2,200)\,\unit\day$},
and sinescale priors of
\mbox{$\mathcal{U}_\text{log}(10^{-2},10^{2})$}.
Our procedure fits one kernel on all quantities, and the latter may have positive or negative correlations; therefore, $A_j,B_j\in\mathbb{R}$. On the other hand, non-gradiented activity indicators have a single parameter, $A_j$. In these cases, we constrained $A_j$ to be non-negative so as to avoid a two-fold degeneracy in \eqref{eq:multidimensional_system}. We used \mbox{$A_j=\mathcal{U}_\text{log}(10^{-3},10^{3})$} and \mbox{$B_j=\mathcal{U}(-10^{3},10^{3})$} for gradiented activity indicators; and $\mbox{$A_j=\mathcal{U}_\text{log}(10^{-3},10^{3})$}$ and \mbox{$B_j=0$} for non-gradiented ones.

\subsection{Inference}

Markov chain Monte Carlo methods can efficiently sample across the high-dimensional parameter spaces of our models. \citet{Skilling2004} noted that while directly sampling from the likelihood function, $\mathcal{L}$, becomes exponentially more expensive, sampling uniformly within a bound \mbox{$\mathcal{L}>k$} is much easier -- and increasing $k$ iteratively upon reaching convergence can be used to evaluate the posterior. This strategy is now known as nested sampling. Its formalism allows it to have:
(i) lower computational complexity;
(ii) the ability to explore multi-modal distributions of arbitrary complex shapes;
(iii) the ability to numerically evaluate the Bayesian evidence, $Z$. The last point, through the Bayes factor, $Z_1/Z_2$, allows for two models to be compared. The Bayes factor describes the relative evidence yielded by data between two models \citep{Morey2016}. We express the Bayes factor in terms of the difference in log-evidence $\Delta\ln Z$ between models. A value of \mbox{$\Delta\ln Z=5$}, for example, would imply that data favours the model at hand by a factor of $e^5\approx 150$. In the literature, a necessary condition for planetary detection is typically $\Delta\ln Z \geq 5$ in favour of the model with a planet. 

We used \mbox{ReactiveNestedSampler}, a nested-sampling integrator provided by \textsc{UltraNest} \citep{ultranest}. In every inference instance of $N_\text{param}$ model parameters, we required $40N_\text{param}$ live points and a region-respecting slice sampler that accepted $4N_\text{param}$ steps until the sample was considered independent.

\section{Analysis}\label{sec:analysis}
\subsection{Features of HARPS-N velocimetry}\label{sec:analysis_features}
The correlation matrix between CCF RVs, LBL RVs and activity indicators reveals two qualities of our dataset (Fig.~\ref{fig:harpsn_correlations}). Firstly, CCF and LBL RVs are very well correlated with one another \mbox{($R= +0.93$)}, and their head-to-head correlation with activity indicators is quite similar \mbox{($|\Delta R|\leq 0.06$)}. We interpret this as a good validation of LBL RVs. Secondly, there is a positive four-way correlation between FWHM, S-index, H$\alpha$ and Na~I ($R\geq 0.43$).

\begin{figure*}
    \centering
    \resizebox{\hsize}{!}{\includegraphics{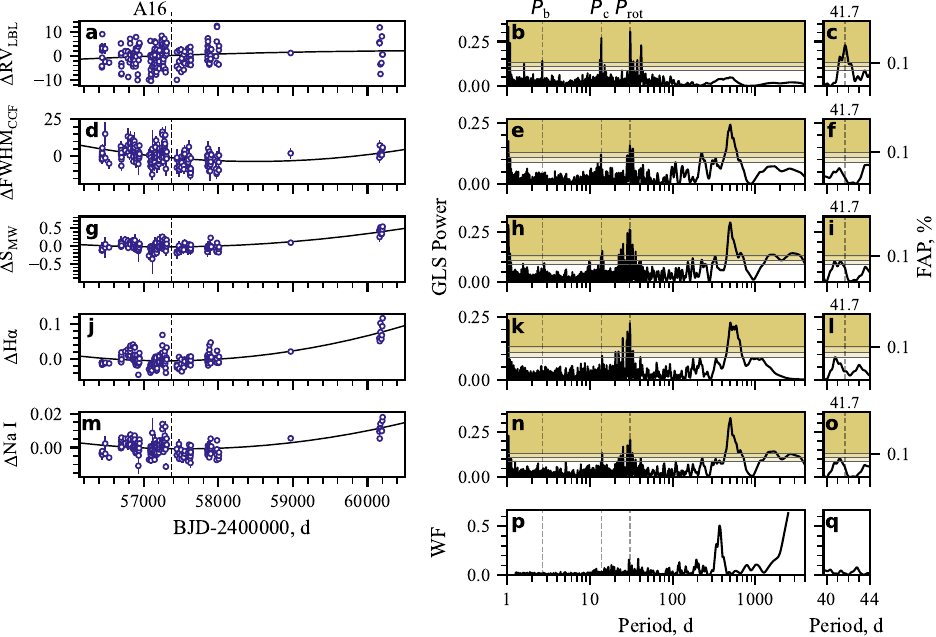}}
    \caption{
    Time series of mean-subtracted:
    (a) LBL RV, in \unit{\metre\per\second};
    (d) CCF FWHM, in \unit{\metre\per\second};
    (g) S-index;
    (j) H$\alpha$;
    (m) Na~I. Measurements are given in hollow blue markers. The temporal coverage of \citetalias{Affer2016} is marked  by a dashed black line. The time series appear to have long-term trends, which we remove with a second-order polynomial fit (solid black line) for the purposes of exploration. (b,e,h,k,n) Associated wide-period GLSPs of LBL RV and activity indicators after subtracting the polynomial fit. Three FAP levels: 10\%, 1\% and 0.1\%, split GLSP ordinates in bands of different colour. We highlight periodicities of the two planets discovered by \citetalias{Affer2016}: $P_\text{b}$ and $P_\text{c}$, as well as the stellar rotational period $P_\text{rot}$ (dashed grey lines). (c,f,i,l,o)~Zoomed-in GLSPs that focus near \SI{41.7}{\day}.
    (p) WF of data.
    (q) Zoomed-in WF that focuses near \SI{41.7}{\day}.
    }
    \label{fig:harpsn_pgrams}
\end{figure*}

Figure~\ref{fig:harpsn_pgrams} shows the generalised Lomb-Scargle periodograms (GLSPs; \citealp{Lomb1976,Scargle1982,Zechmeister2009,VanderPlas2015}) of LBL RV and all activity indicators in a \SIrange{1}{4000}{\day} period interval. There are visible long-term trends in all activity indicators (Fig.~\ref{fig:harpsn_pgrams}d,g,j,m). Second-order polynomial functions handle well these trends, and they help in revealing strong \SIrange{400}{700}{\day} peaks in activity-indicator GLSPs (Fig.~\ref{fig:harpsn_pgrams}e,h,k,n). These peaks appear to preserve their location and strength in the GLSP of time series without a second-order correction, as well in the GLSP of time series excluding points after \SI{2458200}{\julianday}.
This could be indicative of a magnetic cycle -- and as such, it may potentially need to be included in the LTF of our model. We later discuss how our LTF was adapted to handle an inclusion of a cycle (Sect.~\ref{sec:model_comparison}), and to what extent the evidence supports such an inclusion (Sect.~\ref{sec:analysis_cycle}).

 \begin{figure*}
    \centering
    \resizebox{\hsize}{!}{\includegraphics{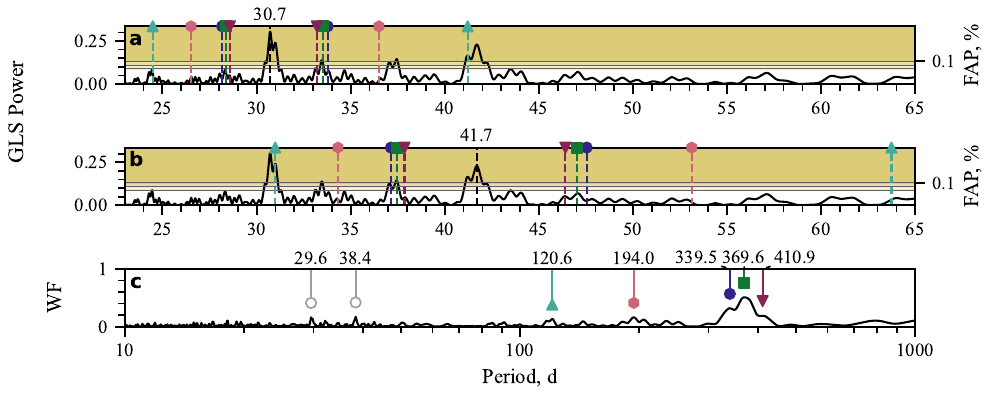}}
    \caption{
    Alias analysis of long-term detrended LBL RV.
    We focus on the aliases of the peaks at: (a)~\SI{30.7}{\day}, the $P_\text{rot}$ peak found in the raw-data GLSP;
    (b)~\SI{41.7}{\day}.
    (c)~The WF reveals five prominent peaks at:
    \SI{120.6}{\day} (upward teal triangle),
    \SI{194.0}{\day} (pink hexagon),
    \SI{339.5}{\day} (blue circle),
    \SI{369.6}{\day} (green square) and
    \SI{410.9}{\day} (downward red triangle).
    Relevant aliases in data periodograms are plotted with dashed lines of a corresponding colour and marker.
    There are other strong peaks in the WF that do not create aliases in the specified period range (hollow grey circles).
    }
    \label{fig:alias_analysis_d}
\end{figure*} 
At the shorter-period end, we identify strong peaks in RV and all activity indicators near \SI{30.7}{\day} (Figs.~\ref{fig:harpsn_pgrams}b,g,l,q,v). We interpret these signals to correspond the stellar rotational period \mbox{$P_\text{rot}=31.8^{+0.6}_{-0.5}\,\unit\day$} measured by \citetalias{Affer2016}. The strongest peak in the RV GLSP lies at $P_\text{rot}$ (Fig.~\ref{fig:harpsn_pgrams}b). 
There, peaks of two other known objects draw attention to themselves:
GJ~3998~b (\SI{2.65}{\day} and its \SI{1}{\day} alias near \SI{1.5}{\day}), as well as
GJ~3998~c (\SI{13.7}{\day}).
However, signals of periodicities close to the planets are observed in some activity indicators: a peak in S-index near GJ~3998~b (\SI{2.80}{\day}; Fig.~\ref{fig:harpsn_pgrams}h), and peaks in all activity indicators near GJ~3998~c (\SI{13.6}{\day} and \SI{14.1}{\day}; Figs.~\ref{fig:harpsn_pgrams}e,h,k,n).

The presence of such activity signatures near planetary periods is problematic. It may suggest stellar-activity origin for these RV signals which had been considered planetary until now. However, activity signals can be independent from the RV signal they neighbour; for example, they can be aliases of activity signatures at other periods. The window function (WF) of our dataset reveals strong signals near \SI{1}{\year} and its harmonics (Fig.~\ref{fig:harpsn_pgrams}p).\footnote{In our work, WFs are the non-centred GLSPs of dataset timestamps with an assigned ordinate value of unity.} We expect that the common harmonics of a true signal, $P_\text{signal}$, and a WF peak at $P_\text{wf}$ come forth as aliases through
\begin{equation}
    P_\text{alias}^{-1}=\left|P_\text{signal}^{-1}\pm P_\text{wf}^{-1}
    \right|
,\end{equation}
as was described in \citet{Dawson2010}. We can therefore identify the most prominent WF peaks, and harmonically combine them with RV peaks to check if the latter are independent from their neighbouring activity peaks. We do this exercise for GJ~3998~b and GJ~3998~c in Appendix~\ref{sec:alias_analysis_bc}, and show that their closest periodicities in activity can be explained by aliasing. We later continue to test possible stellar-activity affiliations to all planetary signals (Sect.~\ref{sec:pseudo_planetary_modelling}).

One more peak catches the eye in the RV GLSP in Fig.~\ref{fig:harpsn_pgrams}b: a \SI{41.7}{\day} signature that we isolate in Fig.~\ref{fig:harpsn_pgrams}c. This signal has been discussed by \citetalias{Affer2016}, who noted its significance not only in RV, but in the S-index and H$\alpha$. The authors considered this peak to be a byproduct of stellar activity, with the signal likely arising due to stellar differential rotation. In our time series, we do not find significant signals at \SI{41.7}{\day} in any activity indicator. However, we do detect strong signals in
S-index (Fig.~\ref{fig:harpsn_pgrams}i) and
Na~I (Fig.~\ref{fig:harpsn_pgrams}o), which both lie near \SI{41.0}{\day} instead. Figure~\ref{fig:alias_analysis_d} shows our test on the dependency between the strong \SI{41.7}{\day} RV signal and $P_\text{rot}$ (measured \SI{30.7}{\day} from the GLSP). These two signals inject aliases close to one another through the same \sfrac{1}{3}\,\unit{\year} alias (Figs.~\ref{fig:alias_analysis_d}a; upward teal triangle). Our \SI{3800}{\day} baseline, however, allows us to resolve the signal and the alias well. We see a clear signal at \SI{41.7}{\day} that cannot come from an alias of the \SI{30.7}{\day} one (i.e. $P_\text{rot}$). Rather, there are two independent signals at \SI{30.7}{\day} and \SI{41.7}{\day}, with each signal having its own family of aliases. There are additional RV structures in the intervals \SIrange{33}{34}{\day} and \SIrange{37}{38}{\day} that may be explained as other aliases of \SI{30.7}{\day} and \SI{41.7}{\day}. Finally, we remark that one of the \sfrac{1}{3}\,\unit{\year} aliases of \SI{30.7}{\day} comes close to \SI{41.0}{\day}, roughly where we observed peaks in S-index, H$\alpha$ and Na~I (Figs.~\ref{fig:harpsn_pgrams}i,l,o).

\begin{figure}
    \centering
    \resizebox{\hsize}{!}{\includegraphics{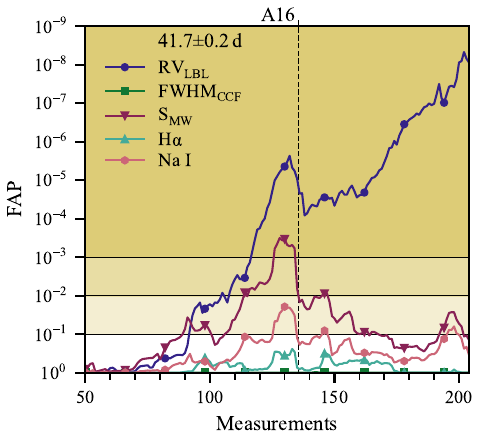}}
    \caption{
    FAP evolution of the strongest peak within \SI{0.2}{\day} of the \SI{41.7}{\day} signal for long-term detrended:
    LBL RV (blue circles),
    CCF FWHM (green squares),
    S-index (downward red triangles),
    H$\alpha$ (upward teal triangles) and
    Na~I (pink hexagons). The temporal coverage of \citetalias{Affer2016} is marked  by a dashed black line.
    }
    \label{fig:harpsn_fap_evolution_d}
\end{figure}

In order to explore the origin and consistency of the \SI{41.7}{\day} RV signal, we constructed stacked GLSPs similar to \citet{Mortier2017}, but in terms of the false-alarm probability (FAP; \citealp{Baluev2008}). Then, we sought the strongest peak within \SI{0.2}{\day} from \SI{41.7}{\day} and computed its FAP, measurement by measurement. Figure~\ref{fig:harpsn_fap_evolution_d} shows this FAP evolution in RV and activity, and can be considered similar to a 1D reduction of Fig.~2 in \citetalias{Affer2016}. Figure~\ref{fig:harpsn_fap_evolution_d} reminds of features that are also seen in \citetalias{Affer2016}: sudden gradual increase in signal from measurements \numrange{70}{80}, with a FAP peak near measurement \num{130}. We marked the boundary of \citetalias{Affer2016} relative to our dataset in terms of time. Until the boundary, the strongest \SIrange{41.5}{41.9}{\day} S-index signal had a <0.1\% FAP. As soon as new measurements enter the dataset, its significance falls to >1\% and even >10\% FAP. This behaviour is mimicked by Na~I, but to a lesser extent. CCF FWHM and Na~I never attain <10\% FAP signals. In contrast, LBL RVs gradually increase in significance from about measurement 70 until the very end.

The significance of \SIrange{41.5}{41.9}{\day} activity signals eventually diminishes. Yet, the strength of these signals until the end of \citetalias{Affer2016} begs an explanation. At that point, the baseline of measurements is almost \SI{900}{\day}, or less than a quarter of our current dataset (\SI{3800}{\day}). With every additional measurement, this baseline increased little by little. Subsequently, the \SI{41.7}{\day} peak and its neighbouring $P_\text{rot}$ alias would become narrower and narrower until they completely separated -- and the $P_\text{rot}$ alias would converge outside the narrow \SIrange{41.5}{41.9}{\day} interval. It is the gradual departure of the $P_\text{rot}$ alias away from the \SIrange{41.5}{41.9}{\day} region that causes a decline in significance for activity signals. GLSPs of broader period intervals (e.g. within \SI{2}{\day} from \SI{41.7}{\day}) show qualitatively the same FAP evolution, but the drop in significance for activity-index signals takes place at a larger number of measurements. This is consistent with expectations -- the $P_\text{rot}$ alias needs a larger baseline to converge away from a wider period interval.

In an idealised case of negligible stellar activity, the FAP of a planetary signal will rise indefinitely in a stacked GLSP by design. In reality, the steady trend of increasing RV significance in Fig.~\ref{fig:harpsn_fap_evolution_d} is marred by a sudden drop between measurements \numrange{135}{170}. GLSPs of broader period intervals preserve the shape of this FAP evolution, suggesting that the strong RV signal is not drifting away from \SI{41.7}{\day}. We found a transfer of powers from the \SI{41.7}{\day} signal to its \SI{1}{\year} alias (Fig.~\ref{fig:harpsn_alias_fap_evolution}; dashed white rectangles). This transfer occurs approximately between measurements \numrange{135}{170}, and would explain why we see such a strong deviation from the expected growth in significance. We constructed stacked GLSPs for GJ~3998~b and GJ~3998~c as well for comparison (Figs. \ref{fig:harpsn_fap_evolution_b} and \ref{fig:harpsn_fap_evolution_c}). The FAP evolution of GJ~3998~c has a similar drop in significance compared to GJ~3998~d.

\subsection{Model comparison}\label{sec:model_comparison}

We adopted the framework in Sect.~\ref{sec:modelling} to build a model grid that probes: (i) different stellar-activity kernels; (ii) planetary systems of different number and type, hereafter `planetary configurations'; and (iii) the possible presence of a sine component in the LTF. For stellar activity, we chose to go with three approximations of the SEP kernel: MEP, ESP with 2 harmonics (hereafter ESP2), and ESP with 3 harmonics (hereafter ESP3). The rank of said kernels increase in this order, and they can be regarded to grow in complexity likewise. In the design of different planetary configurations, we restricted ourselves to cases where inner planets may have circular orbits, while outer planets may have Keplerian orbits -- but not vice versa. We started out by modelling for two to three planets. This resulted in 7 unique configurations: (i) two circular planets; (ii) circular inner and Keplerian outer; (iii) two Keplerian planets; (iv) three circular planets; and so on. In addition to those 7, we included a planet-free configuration and a one-circular configuration, so as to confirm the inner two planets discovered by \citetalias{Affer2016}. This gave 9 planetary configurations in total. In summary, the 3 stellar-activity kernels (MEP, ESP2, ESP3), the 9 planetary configurations and the presence/absence of a sine component in the LTF resulted in a grid of 54 models.

Models were fitted on the conjunction of LBL RV \& CCF FWHM time series. We modelled stellar activity in the multidimensional regime, with FWHM being non-gradiented. We used our default SEP hyperparameter priors (Sect~\ref{sec:parameter_priors}), even though $P_\text{rot}$ had already been constrained in \citetalias{Affer2016}. For the three modelled planets, we used the period priors:
\mbox{$\mathcal{U}\left(2,3\right)\,\unit\day$},
\mbox{$\mathcal{U}\left(13,14\right)\,\unit\day$} and
\mbox{$\mathcal{U}\left(30,50\right)\,\unit\day$}. The priors of the first two planets are informed by \citetalias{Affer2016}, and the third-planet prior tests for the \SI{41.7}{\day} signal, while allowing it to diverge to $P_\text{rot}$. Planetary semi-amplitudes were assigned priors of
\mbox{$k_\text{rv}=\mathcal{U}\left(0,5\right)\,\unit{\metre\per\second}$}; and in Keplerian orbits, they took eccentricity priors of 
\mbox{$e=\mathcal{U}\left(0,1\right)$}. For models with a sine component in the LTF, we used
\begin{equation}
    f_j(t_{i,j}) =
    k_{\text{cyc, }j} \sin\left[\
    \frac{2\pi(t_{i,j}+P_\text{cyc}\,\varphi_{\text{cyc, }j})}{P_\text{cyc}}
    \right] +
    \alpha_{j}t^2_{i,j}+\beta_{j}t_{i,j}+\gamma_{j},
    \label{eq:corrector_function}
\end{equation}
where $k_{\text{cyc, }j}$ and $\varphi_{\text{cyc, }j}$ are the cycle amplitude and cycle phase of quantity $j$, and $P_\text{cyc}$ is the common cycle period. We assigned
amplitude priors of
\mbox{$\mathcal{U}\left(0,40\right)\,\unit{\metre\per\second}$},
phase priors of \mbox{$\mathcal{U}\left(0,1\right)$} and common-period priors of
\mbox{$\mathcal{U}_{\log} \left(200,2000\right)\,\unit\day$}. For models with a sine-free LTF, we set \mbox{$k_{\text{cyc, }j}=0$} and excluded $k_{\text{cyc, }j}$, $\varphi_{\text{cyc, }j}$ and $P_\text{cyc}$ from sampling.

\begin{figure}
    \centering
    \resizebox{\hsize}{!}{\includegraphics{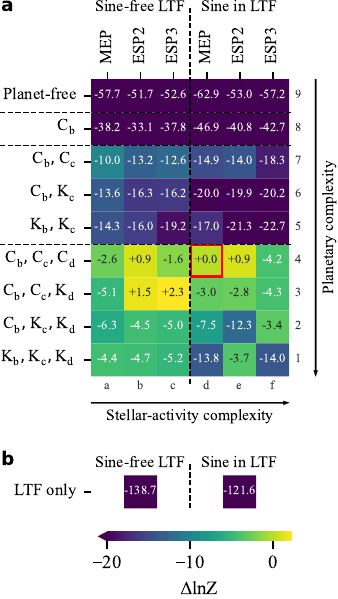}}
    \caption{Bayesian-evidence comparison in our model grid. (a)~Comparison between different planetary configurations and stellar-activity kernels, as well as for a potential presence of a sine component in the LTF (Sect.~\ref{sec:model_comparison}).
    Planetary configurations include planets with circular (C; $e=0$) or Keplerian (K; $e\geq 0$) orbits. For stellar activity, we utilised the MEP kernel and the ESP kernel with 2 or 3 harmonics. The model that we further elect for analysis assumed a sine in the LTF, three circular orbits and a MEP kernel (d4, red border; $\ln Z=-1114.4$).
    We give the Bayesian factor $\Delta\ln Z$ of remaining models relative to model d4.
    (b)~$\Delta\ln Z$ comparison of GP-free planet-free models that only model the LTF. We use this to break the tie between inclusion/exclusion of a sine in the LTF.
    }
    \label{fig:model_table}
\end{figure}

Figure~\ref{fig:model_table} provides with the $\ln Z$ results of our model-grid search. Entries are arranged such that planetary configurations increase in complexity downwards, while stellar-activity kernels increase in complexity rightwards. Dashed lines split the model grid in number of planets (from zero to three); as well as whether a sine term was present in the LTF. We add a chess-like coordinate system at the grid borders, which we later use to refer to a particular model for brevity. Finally, we highlight the most appropriate (`best') model in our analysis: d4, a model with a sine component in the LTF, three circular planets and a MEP stellar-activity kernel. We carry model d4 over to a complete analysis, and we discuss the implications of its results in Sect.~\ref{sec:discussion}.

Later in this work, we compare groups of models that share common features; for example, planet-free models (a9-f9) against one-planet models (a8-f8). We are generally interested in the improvement from a group of less complex models, A, to a group of more complex models, B. We quantify said improvement through two quantities that relate to the Bayes factor. The first quantity is the average improvement in evidence,
\begin{equation}
    (\Delta\ln Z)_\text{avg} = \avg(\ln Z_\text{B}) - \avg(\ln Z_\text{A}),
\end{equation}
with which we report the two groups that formed these averages. The second quantity is the pessimistic improvement in evidence,
\begin{equation}
    (\Delta\ln Z)_\text{pes} = \min(\ln Z_\text{B}) - \max(\ln Z_\text{A}),
\end{equation}
with which we report the models that yielded the $\ln Z$ extrema of each respective group. $(\Delta\ln Z)_\text{pes}$ is defined to favour models from A, so it is naturally more conservative than $(\Delta\ln Z)_\text{avg}$. We use both quantities in tandem, but we regard $(\Delta\ln Z)_\text{pes}$ to better address the true $\ln Z$ uncertainty that has been observed in the literature (e.g. \citealp{Nelson2020}). This comes at the cost of increased false negatives when we try to add new features in models.

\subsection{Evidence of two inner planets}\label{sec:analysis_two}
There are striking differences in evidence between planet-free models and two-planet models (a9-f9 vs. a5-f7; Fig.~\ref{fig:model_table}a). If we consider models with sine-free LTFs,
\mbox{$(\Delta\ln Z)_\text{pes}=32.5$} (b9 vs. c5), implying that data favours two planets by a factor of \mbox{$e^{32.5}>\num{e14}$}. The inclusion of a sine component in the LTF marginally hinders the improvement to a two-planet model -- we report a minimum 
\mbox{$(\Delta\ln Z)_\text{pes}=30.3$} (e9 vs. f5). One-circular models serve as an intermediate point in our analysis, and their $\ln Z$ values fit nicely within our expectations for a smooth increase in evidence. On these grounds, we support the interpretation of \citetalias{Affer2016} and validate the planetary nature of GJ~3998~b and GJ~3998~c. We note strong evidence in favour of circular orbits. For models with sine-free LTFs, and using the two-circular model group for reference, we report the following improvements:
\mbox{$(\Delta\ln Z)_\text{avg}=-3.5$} to a circular inner and a Keplerian outer planet (a7-c7 vs. a6-c6); as well as
\mbox{$(\Delta\ln Z)_\text{avg}=-4.6$} to a two-Keplerian model (a7-c7 vs. a5-c5). For models with a sine component in their LTFs, and using the same group for reference, we report: \mbox{$(\Delta\ln Z)_\text{avg}=-4.3$} (d7-f7 vs. d6-f6); as well as \mbox{$(\Delta\ln Z)_\text{avg}=-4.6$} (d7-f7 vs. d5-f5). \citetalias{Affer2016} published a circular GJ~3998~b and a Keplerian GJ~3998~c, but we point out that the eccentricity posterior of the latter was already compatible with the zero \mbox{($0.049^{+0.094}_{-0.045}$; 2$\sigma$)}. 

\subsection{Evidence for a long-term cycle}\label{sec:analysis_cycle}
Our data supports GJ~3998~b and GJ~3998~c regardless of whether the model LTF contains a sine component (Sect.~\ref{sec:analysis_two}). It is however not evident from the model grid whether data favours a sine component in the LTF on the whole. For this reason, we prepared two intermediate models that simply fit a LTF instead of removing it from the data; consequently, no GPs or planets were included. Figure~\ref{fig:model_table}b compares them head-to-head, aside from the main model grid. The inclusion of a sine component in the LTF increases the evidence by
\mbox{$\Delta\ln Z=17.1$}.

The \mbox{$(\Delta\ln Z)_\text{avg}$} improvements from models with a sine-free LTF to models with a sine in their LTF are:
$-3.7$ for planet-free models (a9-c9 vs. d9-f9);
$-7.1$ for one-circular models (a8-c8 vs. d8-f8);
$-4.1$ for two-planet models (a5-c7 vs. d5-f7);
$-2.8$ for three-planet models (a1-c4 vs. d1-f4).
Unlike the comparison in Fig.~\ref{fig:model_table}b, data do not favour a sine in the LTF as soon as a GP is involved. This suggests that our stellar-activity kernels fit for this long-term trend, which is beyond their intended purpose. It is therefore best to use the results from Fig.~\ref{fig:model_table}b to break the tie between LTFs with and without sines. From now on, we continue to consider only models that include a sine in their LTF (i.e. d1-f9), unless stated otherwise. 

We now turn to the posteriors of the sine components in LTFs. Figure~\ref{fig:model_corrector_period_distro} compares the posterior distributions of the period $P_\text{cyc}$, the RV amplitude $k_\text{cyc, 0}$ and the FWHM amplitude $k_\text{cyc, 1}$ between all relevant models. Distributions are normalised to have equal height for the sake of readability. Posteriors agree well across two- and three-planet models. 
They give an overall impression of a \mbox{$P_\text{cyc}\approx\SI{300}{\day}$} cycle with a well-defined RV amplitude of \SIrange{0.8}{1.4}{\metre\per\second} and a moderately defined FWHM amplitude of \SIrange{0.4}{3}{\metre\per\second}. ASAS-SN Sloan g photometry is characterised by strong peaks near $P_\text{cyc}$ and $2P_\text{cyc}$ as well (Appendix~\ref{sec:photometry_analysis}; Fig.~\ref{fig:asassn_gls}). This may suggest a photometric manifestation of the same cycle. 

Beyond that, the full picture is very different. Planet-free and one-circular models catch the \SI{300}{\day} periodicity, but a \SIrange{500}{600}{\day} mode tends to dominate in the posterior, and sometimes a third mode near \SI{2000}{\day} manifests itself. We also observe differences between kernels -- for example, ESP-kernel $k_\text{cyc, 1}$ posteriors tend to have heavy left tails overall. The latter phenomenon is especially pronounced for ESP3 kernels, for which the mode does not seem to be well separated from the zero. This may come from the higher complexity of the ESP3 kernel, which tries to assign more long-term features to stellar activity. Finally, it may be the data that suggest a poorly constrained FWHM amplitude -- we remind the reader that RV and FWHM measurements have median uncertainties of \SI{1.02}{\metre\per\second} and \SI{3.42}{\metre\per\second}, respectively.

\subsection{Evidence for a third planet}\label{sec:analysis_three}
Three-planet models confidently stand out from their two-planet counterparts. This is true for each stellar-activity kernel taken individually. We report the following \mbox{$(\Delta\ln Z)_\text{avg}$} from two-planet to three-planet models with sines in their LTFs:
11.2 for MEP (d5-d7 vs. d1-d4),
14.0 for ESP2 (e5-e7 vs. e1-e4) and
14.0 for ESP3 (f5-f7 vs. f1-f4). For the same groups, we report \mbox{$(\Delta\ln Z)_\text{pes}$} of 1.0, 1.8, and 4.3, respectively. We can compare two- and three-planet models on a more equal footing by considering models with all-circular and all-Keplerian orbits. For all-circular orbits, the \mbox{$\Delta\ln Z$} improvement is
14.9 for MEP,
14.9 for ESP2 and
14.1 for ESP3.
For all-Keplerian orbits, we report \mbox{$\Delta\ln Z$} of 3.2, 17.6, and 8.7, respectively. All but one of these reported improvements are in principle enough to support a detection in an agnostic manner (\mbox{$\Delta\ln Z \geq 5$}; \citealp{Jeffreys1946}).

To test the numerical stability of third-planet solutions, we took all three planet-models, including ones with sine-free LTFs -- and then individually compared the third-planet orbital-parameter posteriors between themselves. We list the comparison between $P_\text{d}$, $\varphi_\text{d}$, $k_\text{rv, d}$, $e_\text{d}$ and $\omega_\text{d}$ posteriors in Tables
\ref{tab:model_comparison_planetd_period},
\ref{tab:model_comparison_planetd_phase},
\ref{tab:model_comparison_planetd_krv},
\ref{tab:model_comparison_planetd_e}, and
\ref{tab:model_comparison_planetd_w}, respectively.
All models successfully recover an RV signature of period \SIrange{41.77}{41.79}{\day} and of semi-amplitude \SIrange{1.74}{1.91}{\metre\per\second}. Posteriors agree in uncertainty for each orbital parameter, and for any model pair.

\begin{figure}
    \centering
    \resizebox{\hsize}{!}{\includegraphics{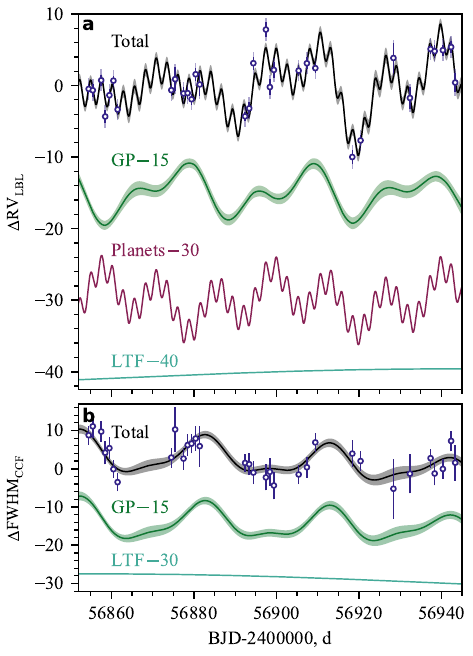}}
    \caption{
    Raw time series (blue points) against our best model (d4; solid lines) in a selected temporal region. Data points come with two error bars: one assigned by the instrument (blue), and another that includes the model jitter (grey; Sect.~\ref{sec:modelling_offsets_jitters}).
    (a) LBL RV in \unit{\metre\per\second}, where the full model (black) is split into the stellar-activity GP component (green), the three-planet component (red) and the LTF (teal).
    (b) CCF FWHM in \unit{\metre\per\second}, where the full model (black) is split into the stellar-activity GP component (green) and the LTF (teal). Every component is offset by an arbitrary amount, labelled in the plots. The $1\sigma$ confidence intervals of the GP are given in shaded bands.
    }
    \label{fig:bestmodel_timeseries}
\end{figure}

Among three-planet models with sines in their LTFs, the three-circular MEP model is the simplest, both in terms of planetary- and in stellar-activity complexity. Some other models fare better in evidence, but not by much \mbox{($\Delta\ln Z\leq 2.3$)}. We therefore select d4 as our best-fitting model, or best model for short. Figure~\ref{fig:bestmodel_timeseries} highlights the agreement between model d4 and data in a selected dense interval with 32 clustered data points. As d4 models for three planets, the model fit time series expectedly shows much more complexity in RV. Nevertheless, data appear to be fitted well within uncertainty in both quantities. The temporal sampling of our measurements is non-trivial, as is hinted at by the WF (Fig.~\ref{fig:harpsn_pgrams}p). There are clusters of more than ten measurements within \SI{20}{\day} (e.g. \SIrange{2457140}{2457160}{\julianday}), but also lone measurements that are isolated in time by several tens of days (e.g. near \SI{2456750}{\julianday}). We inspected the model fit in the full time series, and the MEP kernel appears to have no problem with modelling activity on short and long timescales.

 \begin{figure}
    \centering
    \resizebox{\hsize}{!}{\includegraphics{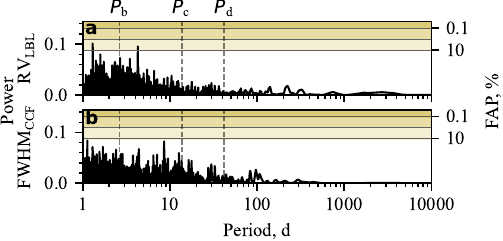}}
    \caption{
    GLSPs of the residual time series of our best model (d4), accounting for the model jitter, for: (a) LBL RVs, (b) CCF FWHMs.
    }
    \label{fig:bestmodel_pgram}
\end{figure} 
\begin{figure*}
    \centering
    \resizebox{\hsize}{!}{\includegraphics{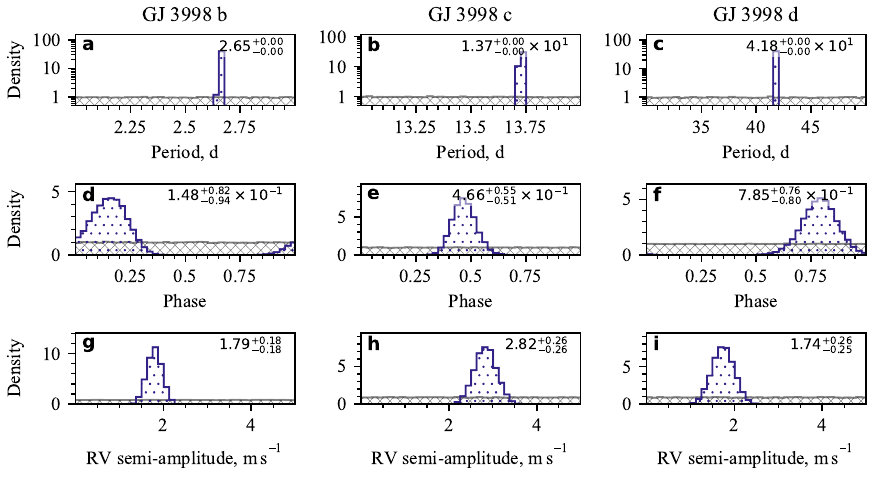}}
    \caption{Prior (grey, cross-hatched) and posterior (blue, dot-hatched) distributions of planetary parameters in our best model (d4). Uncertainties in our posteriors reflect the 16\textsuperscript{th} and the 84\textsuperscript{th} percentiles.
    }
    \label{fig:bestmodel_planet_posteriors}
\end{figure*} \begin{figure*}
    \centering
    \resizebox{\hsize}{!}{\includegraphics{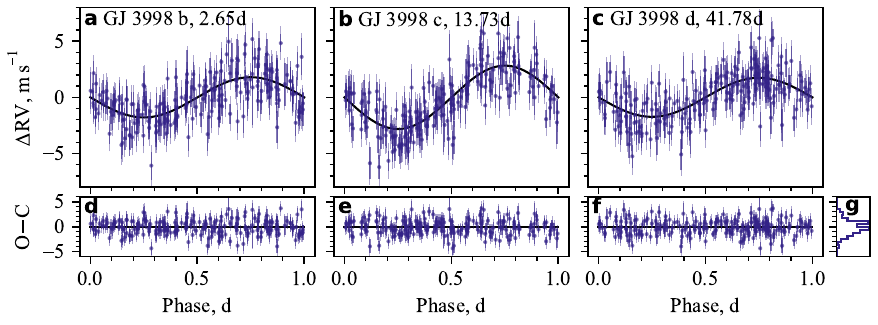}}
    \caption{
    (a,b,c)~Activity-corrected, phase-folded time series (blue error bars) against the planetary fit of our best model (d4; solid black line). (d,e,f)~Residual time series after the planetary fit. (g)~Distribution of residuals (solid blue line).
    }
    \label{fig:bestmodel_planets}
\end{figure*} 
Figure \ref{fig:bestmodel_pgram} supplies with the RV and FWHM GLSPs of the residual time series, having accounted for the model jitter. Both periodograms show no significant periodicities except a 3\% FAP RV signal at \SI{1.30}{\day} and a 5\% FAP RV signal at \SI{4.28}{\day}. At the same time, they do not hint of overfitting. Figure \ref{fig:bestmodel_planet_posteriors} displays the parameter posteriors of the three planets, and Fig.~\ref{fig:bestmodel_planets} shows their isolated RV signals from modelled trends and stellar-activity.
Data agrees well with each modelled sinusoidal signal, including the novel GJ~3998~d. We observe ordinary residual distributions and no apparent trends with phase.

Table \ref{tab:bestmodel_posterior} compares our posteriors with the published results by \citetalias{Affer2016}. Although not all common parameters could be compared between \citetalias{Affer2016} and our work, we find reasonable agreement for the ones that do. Our RV jitter
\mbox{$J_{0,0}=1.29^{+0.15}_{-0.14}\,\unit{\metre\per\second}$} agrees with \citetalias{Affer2016} ($1.19^{+0.11}_{-0.14}\,\unit{\metre\per\second}$), but is marginally larger. This could be indicative of the lack of instrumental-drift correction (Sect.~\ref{sec:harpsn_velocimetry}). Our stellar-rotation period
\mbox{$P_\text{rot}=30.2\pm 0.3\,\unit\day$}
somewhat agrees with \citetalias{Affer2016} ($31.8^{+0.6}_{-0.5}\,\unit{\day}$). The posteriors of GJ~3998~b and GJ~3998~c are in excellent agreement with \citetalias{Affer2016}, with one exception: our inferior-conjunction ephemerides have uncertainties that are three to seven times as large. This may come from the choice of ephemeris sampling: we sampled in normalised phase, whereas \citetalias{Affer2016} did so in epoch directly. Our third-planet posteriors appear well defined in model d4. The RV semi-amplitude of this new planet ($1.74^{+0.26}_{-0.25}\,\unit{\metre\per\second}$) stands nearly $7\sigma$ away from the zero. Its phase is well defined, and has a similar absolute uncertainty compared to the other two planets.

In addition to inferior-conjunction ephemerides, we derived three other quantities in Table~\ref{tab:bestmodel_posterior}: semi-major axes $a$, minimum planetary masses $m\sin i$ and mean insolation fluxes $\Phi$. The first two quantities necessitate knowledge of the stellar mass $M_\star$, while the third quantity requires $M_\star$ and the stellar luminosity $L_\star$. Through stellar parameters from Table~\ref{tab:stellar_parameters}, we get compatible results for the semi-major axes and the minimum masses of the inner two planets. For the third planet, we obtain
\mbox{$m_\text{d}\sin i_\text{d} = 6.07^{+1.00}_{-0.96}\,\unit\earthmass$}. We report
\mbox{$\Phi_\text{b}=48^{+11}_{-9}\,\unit\earthflux$}, 
\mbox{$\Phi_\text{c}=5.4^{+1.3}_{-1.0}\,\unit\earthflux$} and
\mbox{$\Phi_\text{d}=1.2^{+0.3}_{-0.2}\,\unit\earthflux$}. We discuss the implications of these results in Sect.~\ref{sec:planetary_system}.

\begin{figure}
    \centering
    \resizebox{\hsize}{!}{\includegraphics{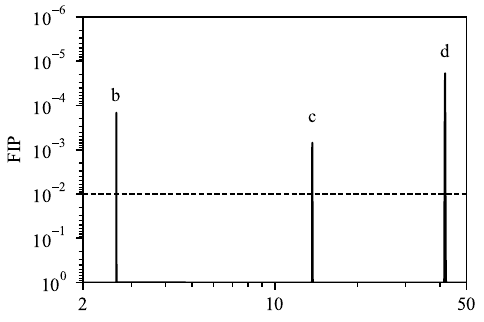}}
    \caption{FIP-period plot of our best model (d4; solid black line) against the \num{e-2} threshold suggested by \citet{Hara2022} (dashed black line).}
    \label{fig:model_fip}
\end{figure}

We assessed the significance of GJ~3998~b, GJ~3998~c and GJ~3998~d following \citet{Hara2022}, hereafter \citetalias{Hara2022}. Their formalism uses the posterior distribution of an inference run to compute the probability of having no planets within a given element in the orbital-frequency space. \citetalias{Hara2022} refers to this metric as the false inclusion probability (FIP). We re-ran model d4 with the following modification in priors:
$P_\text{b}, P_\text{c}, P_\text{d} =\mathcal{U}(2,50)\,\unit\day$. We did so in order to encompass the full period range \SIrange{2}{50}{\day}, which had not been fully covered by our individual priors. The formalism by \citetalias{Hara2022} works in Bayesian-evidence terms, at a stage where the parameter space has been already marginalised. Therefore, having identical $P_\text{b}, P_\text{c}, P_\text{d}$ priors should not perturb the FIP framework. We computed the FIP between \SIrange{2}{50}{\day} for an angular-frequency step \mbox{$\Delta\omega=2\pi/5T_\text{span}$}, where $T_\text{span}$ is the observation span. Figure~\ref{fig:model_fip} displays our FIP calculation as a function of orbital period. Our new discovery, GJ~3998~d, is assigned a FIP of around \num{2e-5}, which passes the \num{e-2} threshold suggested by \citetalias{Hara2022}.

\subsubsection{Modelling with RV and other activity indicators}\label{sec:rv_activity_indicators}

We ran model d4 on RV in conjunction with other activity indicators: RV\&S-index, RV\&H$\alpha$, and RV\&Na~I. If the \SI{41.7}{\day} signal comes from stellar activity, then model d4 would fail to describe the observed time series, or would converge to a different result. We ran these models again in the multi-dimensional regime, with activity indicators being non-gradiented in the setup, similar to FWHM. We used the same priors as in model d4, but with one alteration: we used $J_{0,1}=\mathcal{U}_{\log}\left(10^{-4},1\right)$ for H$\alpha$ and Na~I to account for their small absolute uncertainties. Finally, we repeated model d4, but using CCF RVs instead of LBL RVs. This was done to check if the detection of a third planet could have been made through CCF data alone.

Table \ref{tab:planet_recovery_indicators} compares the posteriors of model d4 against those run variations. We could only find four disagreements between comparable parameters. The first two disagreements come from the CCF RV \& CCF FWHM model:
(i)~the LTF sine period \mbox{$P_\text{cyc}$} tries to converge near its \SI{2000}{\day} prior boundary,
(ii)~$P_\text{d}$ contains two secondary modes approximately at \SI{31.1}{\day} and \SI{31.4}{\day}, close to the \sfrac{1}{3}\,\unit{\year} alias of the \SI{41.7}{\day} signal.
In third place comes the very unconstrained \mbox{$P_\text{cyc}=1384^{+502}_{-945}\,\unit\day$} of the LBL RV \& Na~I model.
Finally, the LTF of the LBL RV \& S-index model has flat phase posteriors.

Among these four disagreements, only the second relates to a planetary parameter. It is not fundamental and it takes place in a dataset where RV is of known lower precision. The CCF RV \& CCF FWHM model has posteriors that are unsurprisingly less constrained than the ones from model d4. Nevertheless, GJ~3998~d is detected even with CCF RVs -- we measure an RV semi-amplitude of $1.52^{+0.41}_{-0.45}\,\unit{\metre\per\second}$, more than $3\sigma$ away from the zero.

\subsubsection{Modelling with RVs from other pipelines}\label{sec:serval}
We used SERVAL \citep{serval} to compute a third RV data source, with which we can validate CCF and LBL RVs. We reduced all available raw HARPS-N spectra through SERVAL, and we took from there only RVs and its associated uncertainties. Then, we crossmatched our dataset with SERVAL RVs by time, thereby rejecting the same outliers. This query returned 204 datapoints with a median temporal spacing of \SI{1.0}{\day}. \mbox{SERVAL} RVs have an rms of \SI{4.25}{\metre\per\second}, which approaches LBL RV rms (\SI{4.28}{\metre\per\second}). These new RV measurements are less precise, with a median uncertainty of \SI{1.19}{\metre\per\second} (against \SI{1.02}{\metre\per\second} in LBL RVs). We ran a duplicate of model d4, but on SERVAL RVs \& CCF FWHM data. We found no differences from our d4 results (Fig.~\ref{fig:bestmodel_posteriors}, Table~\ref{tab:bestmodel_posterior}) that warrant discussion.

\subsubsection{Modelling activity signals near planetary periods}\label{sec:pseudo_planetary_modelling}

Planets manifest themselves in RV only, not in activity indicators. Therefore, strong activity signals near orbital periods of planets should prompt scrutiny. This concept has been fundamental for the validation of detections since the early days of exoplanetary discovery (e.g. \citealp{Queloz2001}), and is still being used to refute planetary candidates (e.g. \citealp{Robertson2015}). In fact, GJ~3998 has already been subject of similar analysis. Recently, \citet{Dodson-Robinson2022} analysed the historical \citetalias{Affer2016} data and shared concerns that the signals attributed to GJ~3998~b and GJ~3998~c may come from stellar activity.

Not all cases, however, are easy to resolve. For example, the stellar rotational period of GJ~3998 is
\mbox{$P_\text{rot}=30.2\pm 0.3\,\unit\day$}, and the orbital period of GJ~3998~c is
\mbox{$P_\text{c}=13.727^{+0.003}_{-0.004}\,\unit\day$}; in other words, near $P_\text{rot}/2$ (Table~\ref{tab:bestmodel_posterior}). Figure~\ref{fig:harpsn_pgrams} was earlier discussed to reveal strong peaks in activity near $P_\text{c}$. This would either imply that: (a) the \SI{13.7}{\day} RV signal is indeed a planet and the \SI{13.7}{\day} activity signals are harmonics of $P_\text{rot}$; or that (b) RV and activity signals come from the same stellar-activity component, which somehow escapes our modelling. Because case (b) would invalidate a planet, we found it necessary to devise a general test for all planets in the system.

Suppose there is an excess signal at period $P$ in RV and activity, which our modelling fails to address in all physical quantities. Therefore, a model that accounts for stellar activity and one planet should erroneously guide the planetary period to $P$. However, because the excess signal manifests in activity too, if we fit a planet-like component in activity instead of RV, the period of this component should also be erroneously guided to $P$. We can therefore fit variants of model d4 with three sine terms in a chosen activity indicator, and check if the posteriors of those terms compare to the three planets in model d4. If posteriors show similarity for any planet, case (b) holds and that planet must be invalidated.

\begin{figure*}
    \centering
    \resizebox{\hsize}{!}{\includegraphics{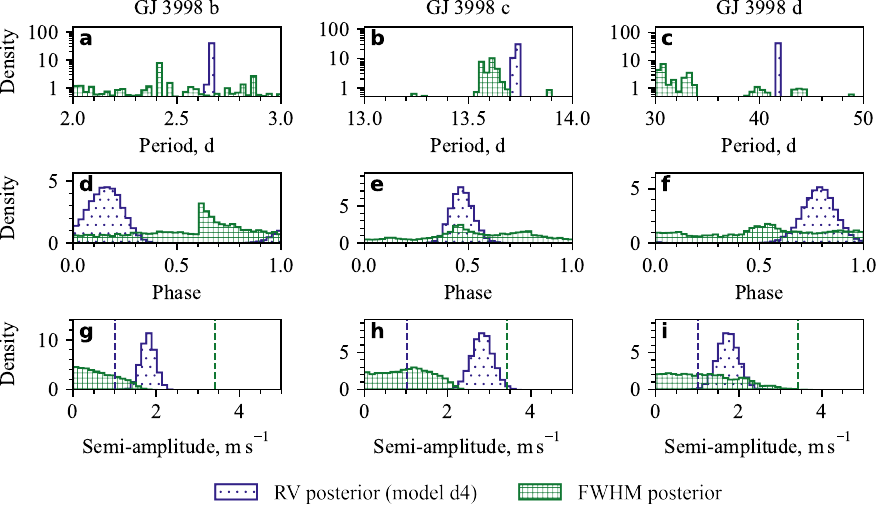}}
    \caption{
    Planetary-parameter posteriors of our best model (d4; Fig.~\ref{fig:bestmodel_planet_posteriors}; blue dot-hatched) against the posteriors from our FWHM model with three activity sine terms (green square-hatched). Parameters share period- and phase priors across models. We use semi-amplitude panels (g,h,i) to show  the median uncertainty of RV and FWHM measurements (dashed lines in respective colour).
    }
    \label{fig:bestmodel_fwhmmodel_comparison}
\end{figure*}

\begin{figure*}
    \centering
    \resizebox{\hsize}{!}{\includegraphics{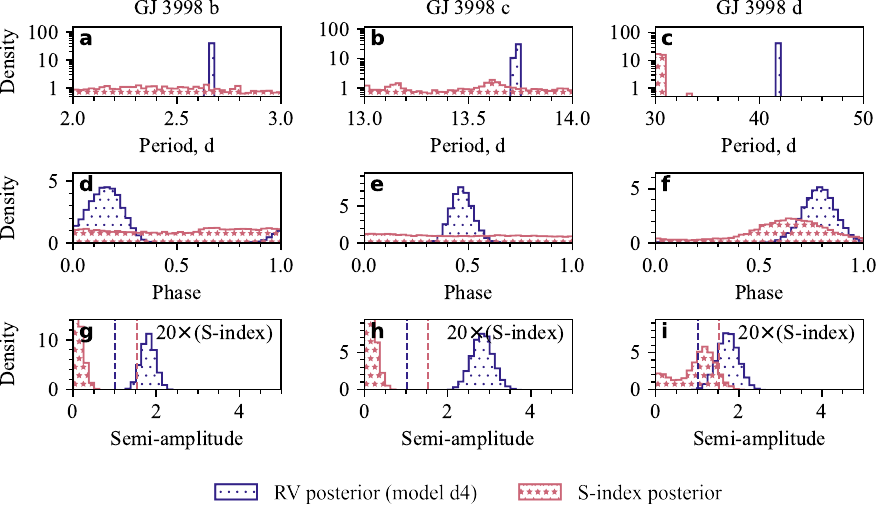}}
    \caption{
    Identical to Fig.~\ref{fig:bestmodel_fwhmmodel_comparison} but making a comparison between our best model (d4; Fig.~\ref{fig:bestmodel_planet_posteriors}; blue dot-hatched) and our S-index model with three activity sine terms (pink star-hatched). To ease comparison, semi-amplitudes and the median uncertainty in Na~I were scaled by \num{20}.
    }
    \label{fig:bestmodel_sindmodel_comparison}
\end{figure*}

\begin{figure*}
    \centering
    \resizebox{\hsize}{!}{\includegraphics{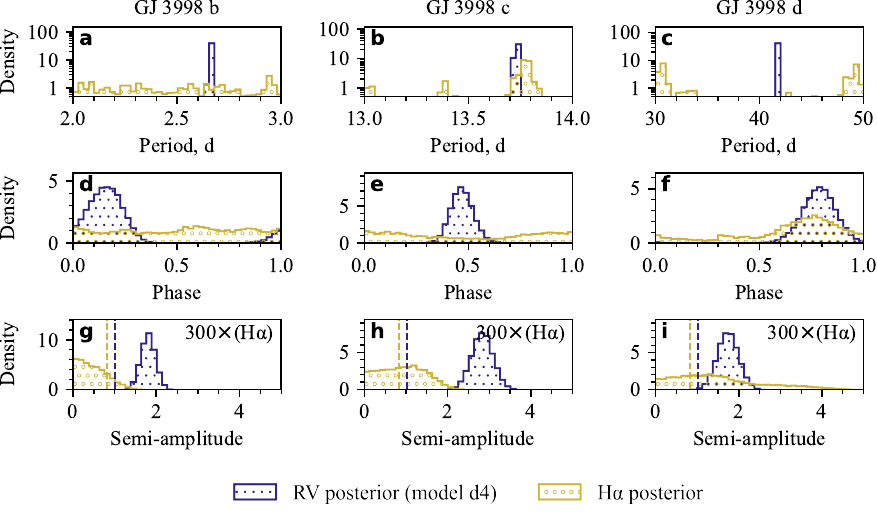}}
    \caption{
    Identical to Fig.~\ref{fig:bestmodel_fwhmmodel_comparison} but making a comparison between our best model (d4; Fig.~\ref{fig:bestmodel_planet_posteriors}; blue dot-hatched) and our H$\alpha$ model with three activity sine terms (yellow circle-hatched). To ease comparison, semi-amplitudes and the median uncertainty in H$\alpha$ were scaled by \num{300}.
    }
    \label{fig:bestmodel_halpmodel_comparison}
\end{figure*}

\begin{figure*}
    \centering
    \resizebox{\hsize}{!}{\includegraphics{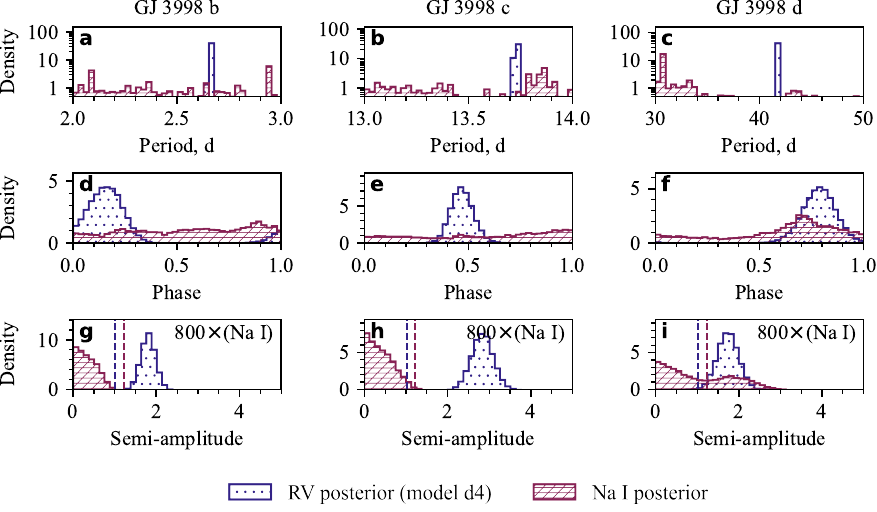}}
    \caption{
    Identical to Fig.~\ref{fig:bestmodel_fwhmmodel_comparison} but making a comparison between our best model (d4; Fig.~\ref{fig:bestmodel_planet_posteriors}; blue dot-hatched) and Na~I model with three activity sine terms (red rhomboid-hatched). To ease comparison, semi-amplitudes and the median uncertainty in Na~I were scaled by \num{800}.
    }
    \label{fig:bestmodel_naimodel_comparison}
\end{figure*}

We searched for three sinusoidal signals in variants of model d4 that were individually fit on each activity-indicator time series: on FWHM data only, on S-index data only and so on.
We used the same priors as in model d4, with the same adjustments as in Sect.~\ref{sec:rv_activity_indicators}:
\mbox{$J_{0,0}=\mathcal{U}_{\log}\left(10^{-4},1\right)$}
for H$\alpha$ and Na~I to handle the scale of their uncertainties.
Figures 
\ref{fig:bestmodel_fwhmmodel_comparison},
\ref{fig:bestmodel_sindmodel_comparison},
\ref{fig:bestmodel_halpmodel_comparison}, and
\ref{fig:bestmodel_naimodel_comparison}
compare the three-planet posteriors between model d4 and models in: FWHM, S-index, H$\alpha$, and Na~I, respectively. None of these comparisons show evidence to suggest a stellar-activity origin for any of the three planets. A more thorough discussion on the results follows.

    
A baseline proxy for the significance of a signal is its semi-amplitude posterior relative to the median uncertainty of the data. We remind the reader that all three planets in model d4 have RV semi-amplitudes that are well separated from the median RV uncertainty (\SI{1.02}{\metre\per\second}; dashed blue lines in g,h,i panels of Figs.~\ref{fig:bestmodel_fwhmmodel_comparison}-\ref{fig:bestmodel_naimodel_comparison}). Some of the fitted activity sine terms have inklings of modes in semi-amplitude near the median uncertainty -- but they all have other clear disagreements with d4 planet posteriors. These sine terms were designed to seek activity signals in the priors of:
(i)~GJ~3998~d in S-index, Fig.~\ref{fig:bestmodel_sindmodel_comparison}i, which converges to a period between \SIrange{30}{31}{\day}, close to $P_\text{rot}$;
(ii)~GJ~3998~c in H$\alpha$, Fig.~\ref{fig:bestmodel_halpmodel_comparison}h, which converges to a weakly defined period and phase; (iii)~GJ~3998~d in H$\alpha$, Fig.~\ref{fig:bestmodel_halpmodel_comparison}i, which has two period modes: \SIrange{30}{31}{\day} and \SIrange{48}{50}{\day}, both far away from $P_\text{d}$;
(iv)~GJ~3998~d in Na~I, Fig.~\ref{fig:bestmodel_naimodel_comparison}i, which converges to a period between \SIrange{30}{31}{\day}, close to $P_\text{rot}$.

On the whole, we do observe a few faint structures in the posteriors of the fitted sine terms in activity. These structures may indeed come from residual activity that had not been picked up by the stellar-activity kernel, or from the specific WF of our data -- but none of them come close to the well-constrained planetary parameters of model d4. On these grounds, we conclude that this test validates the planetary nature of GJ~3998~b, GJ~3998~c, and GJ~3998~d.

\section{Discussion}\label{sec:discussion}
\subsection{The planetary system of GJ~3998}\label{sec:planetary_system}
We confirm the existence of planets GJ~3998~b and GJ~3998~c that were discovered by \citetalias{Affer2016}. Additionally, we assign the \SI{41.7}{\day} RV signal to a third planet, GJ~3998~d. We constrain the orbital periods of GJ~3998~b, GJ~3998~c and GJ~3998~d to
$2.65033^{+(22)}_{-(19)}$\,\unit\day,\footnote{We report $P_b$ uncertainties in parentheses, to the order of the least significant figure of the nominal value.}
$13.727^{+0.003}_{-0.004}$\,\unit\day, and
\mbox{$41.78\pm 0.05\,\unit\day$}, respectively (Table~\ref{tab:planet_summary}).
These planets have minimum masses of
$2.50^{+0.30}_{-0.29}$\,\unit\earthmass,
$6.82^{+0.78}_{-0.75}$\,\unit\earthmass, and
$6.07^{+1.00}_{-0.96}$\,\unit\earthmass, respectively.
All planets have well-defined posteriors (Tab.~\ref{tab:planet_summary}, Fig.~\ref{fig:bestmodel_planet_posteriors}), and HARPS-N data matches nicely our best-model fit at their periods (Fig. \ref{fig:bestmodel_planets}). Our best-model residual time series reveal no significant signals in RV nor FWHM (Fig.~\ref{fig:bestmodel_pgram}). All planets likely have minute eccentricities, but our current data and modelling do not permit us to provide lower limits.

\begin{figure*}
    \centering
    \resizebox{\hsize}{!}{\includegraphics{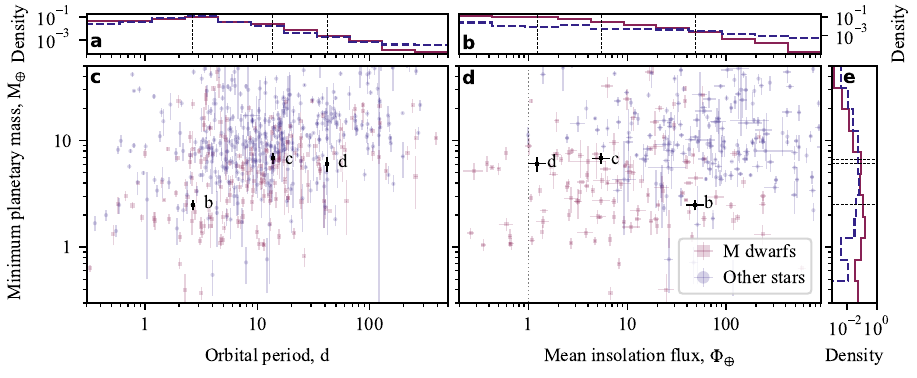}}
    \caption{The physical parameters of the GJ~3998 planetary system against the literature. Centre: Mass-flux and minimum mass-period diagrams of exoplanets with well-established masses and periods from \citet{nasaexoplanettable} (red squares for M-dwarf hosts, blue circles otherwise). Against those, we plot the derived posteriors of our best model (d4; black error bars) using stellar properties from \citet{Maldonado2020}. Reported uncertainties reflect the 16\textsuperscript{th} and the 84\textsuperscript{th} percentiles of our posteriors. Top and right sides: one-dimensional histograms of well-established planets from \citet{nasaexoplanettable} (solid red lines for M-dwarf hosts, dashed blue lines otherwise). The locations of GJ~3998~b, GJ~3998~c and GJ~3998~d are marked by dashed black lines.
    }
    \label{fig:planet_distributions}
\end{figure*}

\begin{table}
    \centering
    \caption{Planetary parameters from our best model (d4; Table~\ref{tab:bestmodel_posterior}).
    }
    \label{tab:planet_summary}
    
    \begin{tabular}{lcl}
    \hline \hline
    Parameter symbol &
    Unit &
    This work \\\hline
    \textbf{GJ~3998~b}\\
    $P_\text{b}$ &
    \unit\day &
    $2.65033^{+(22)}_{-(19)}$ \\
    $\varphi_\text{b}$ &
    - &
    $0.148^{+0.082}_{-0.094}$ \\
    $\varepsilon_\text{b}$ &
    \unit\julianday$-\num{2400000}$ &
    $60202.964^{+0.249}_{-0.216}$ \\
    $k_\text{rv, b}$ &
    \unit{\metre\per\second} &
    $1.79\pm 0.18$ \\
    $e_\text{b}$ &
    - &
    0 \\
    $a_\text{b}$ &
    \unit\au &
    $0.030\pm 0.001$ \\
    $m_\text{b}\sin i_\text{b}$ &
    \unit\earthmass &
    $2.50^{+0.30}_{-0.29}$ \\
    $\Phi_\text{b}$ &
    \unit\earthflux &
    $48^{+11}_{-9}$ \\
    \textbf{GJ~3998~c}\\
    $P_\text{c}$ &
    \unit\day &
    $13.727^{+0.003}_{-0.004}$ \\
    $\varphi_\text{c}$ &
    - &
    $0.466^{+0.055}_{-0.051}$ \\
    $\varepsilon_\text{c}$ &
    \unit\julianday$-\num{2400000}$ &
    $60196.96^{+0.70}_{-0.75}$ \\
    $k_\text{rv, c}$ &
    \unit{\metre\per\second} &
    $2.86\pm 0.26$ \\
    $e_\text{c}$ &
    - &
    0 \\
    $a_\text{c}$ &
    \unit\au &
    $0.090\pm 0.003$ \\
    $m_\text{c}\sin i_\text{c}$ &
    \unit\earthmass &
    $6.82^{+0.78}_{-0.75}$ \\
    $\Phi_\text{c}$ &
    \unit\earthflux &
    $5.4^{+1.3}_{-1.0}$ \\
    \textbf{GJ~3998~d}\\
    $P_\text{d}$ &
    \unit\day &
    $41.78\pm 0.05$ \\
    $\varphi_\text{d}$ &
    - &
    $0.785^{+0.076}_{-0.080}$ \\
    $\varepsilon_\text{d}$ &
    \unit\julianday$-\num{2400000}$ &
    $60170.55^{+3.29}_{-3.15}$ \\
    $k_\text{rv, d}$ &
    \unit{\metre\per\second} &
    $1.74^{+0.26}_{-0.25}$ \\
    $e_\text{d}$ &
    - &
    0 \\
    $a_\text{d}$ &
    \unit\au &
    $0.189\pm 0.006$ \\
    $m_\text{d}\sin i_\text{d}$ &
    \unit\earthmass &
    $6.07^{+1.00}_{-0.96}$ \\
    $\Phi_\text{d}$ &
    \unit\earthflux &
    $1.2^{+0.3}_{-0.2}$ \\[1ex] \hline
    \end{tabular}
    \tablefoot{
    Reported uncertainties reflect the 16\textsuperscript{th} and the 84\textsuperscript{th} percentiles. $P_\text{b}$ uncertainties are given in parentheses, to the order of the least significant figure of the nominal value.
    }
\end{table} 
We computed the mass-period and the mass-flux relationships of known exoplanets that have had their minimum masses measured independently, and that have relative mass uncertainties smaller than one third \citep{nasaexoplanettable}. Figure~\ref{fig:planet_distributions} presents said diagram for
the period range \SIrange{0.3}{500}{\day},
the flux range \SIrange{0.2}{900}{\earthflux} and
the mass range \SIrange{0.3}{50}{\earthmass} against our posteriors of the three planets in the GJ~3998 system.
With regards to the mass-period diagram, GJ~3998~c and GJ~3998~d lie in the bulk of discovered planets. GJ~3998~b somewhat strays from the main distribution (Fig.~\ref{fig:planet_distributions}c).
It is in the mass-flux diagram, however, where our three planets stand out -- the minimum masses of all put them in less explored fields in the mass-flux space (Fig.~\ref{fig:planet_distributions}d). This is especially the case for GJ~3998~d, which turns out to be one of the few known planets that receive an Earth-like flux. In terms of the exoplanetary statistics of similar-mass stellar hosts ($\pm\SI{0.1}{\solarmass}$), GJ~3998~d would be the second planet of Earth-like flux after GJ~3293~b \citep{Astudillo-Defru2015}.

\begin{figure}
    \centering
    \resizebox{\hsize}{!}{\includegraphics{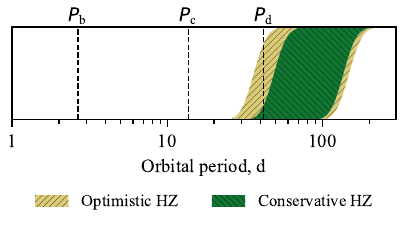}}
    \caption{
    Optimistic (yellow, north-east hatching) and conservative (green, north-west hatching) \SI{5}{\earthmass} HZs from \citet{Kopparapu2014} against orbital periods of GJ~3998~b, GJ~3998~c and GJ~3998~d (vertical dashed lines). We show \num{e5} separate HZ estimates that account for the uncertainty on GJ~3998 stellar parameters by \citet{Maldonado2020}.
    }
    \label{fig:model_habitability}
\end{figure}

The favourable flux of GJ~3998~d raises questions about its potential habitability. A traditional choice to assess the habitability of a planet is through \citet{Kopparapu2014}. Their work provides the runaway- and maximum-greenhouse flux limits, which we used to define a conservative HZ (cHZ). In the same fashion, the authors also provide recent-Venus and early-Mars flux limits, which we take as boundaries of an optimistic HZ (oHZ). We used $M_\star, \log_{10}L, T_\text{eff}$ from  Table~\ref{tab:stellar_parameters} and generated \num{e5} estimations of \SI{5}{\earthmass} cHZs and oHZs. Then, we sorted those estimations by orbital period and plotted them against the best-model periods of our three planets. Figure~\ref{fig:model_habitability} displays the results of this exercise. Our measured $P_\text{d}$ crosses the cHZ, the oHZ, and the uninhabitable zone (uHZ). The length of the $P_\text{d}$ line in each zone is proportional to the probability of said planet residing in said zone. Through our HZ estimations, we report the following probabilities of $P_\text{d}$ belonging in the cHZ, the oHZ, and the uHZ: 17\%, 67\%, and 16\%, respectively. We stress that our form of assessment assumes a rocky-planet composition for GJ~3998~d. This would imply a true mass of \SI{5}{\earthmass} at most. Having assumed a random orbit orientation, the probability of this mass inequality is 5\%.

\subsection{Detection limits for additional companions}
We assessed our detection limits for additional planetary companions with a procedure similar to \citet{SuarezMascareno2023}. We injected a random planetary-sine signal as is described in \eqref{eq:plasine_subtraction}, following the distributions
\mbox{$P=\mathcal{U}_{\log}(1,1000)\,\unit\day$},
\mbox{$k_\text{rv}=\mathcal{U}_{\log}(0.1,5)\,\unit{\metre\per\second}$}, and
\mbox{$\varphi=\mathcal{U}(0,1)$}.
Then, we used the median posterior of our best model (d4; Table~\ref{tab:bestmodel_posterior}) to fit d4 onto the modified time series. The residual time series after this fit should be void of long- and short-term stellar activity, as well as of RV signals from the three planets -- but they may or may not contain the injected signal. We then constructed a GLSP on the residual time series and computed the FAP of the strongest peak in the interval \SIrange{1}{1000}{\day} of 1000 log-spaced points.\footnote{This step includes an RV-jitter term that also comes from model d4.} We repeated this operation until we obtained the FAPs of \num{2e6} random planetary-sine signals.

\begin{figure}
    \centering
    \resizebox{\hsize}{!}{\includegraphics{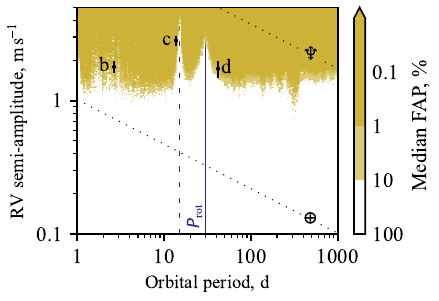}}
    \caption{
    Detection-limit grid of \num{2.4e6} random planetary-sine signals of RV amplitude \SIrange{0.1}{5}{\metre\per\second} and of period \SIrange{1}{1000}{\day}. The grid is formed by \numproduct{200x200} log-spaced bins. Stellar rotation complicates detection considerably for orbital periods near $P_\text{rot}$ (solid blue line) and $P_\text{rot}/2$ (dashed blue line). We display our best-model solutions (d4; black error bars) alongside solutions of Neptune- and Earth-mass planets (dotted black lines). Reported uncertainties reflect the 16\textsuperscript{th} and the 84\textsuperscript{th} percentiles.
    }
    \label{fig:model_detectability}
\end{figure}

Figure~\ref{fig:model_detectability} displays the median FAP of all injected signals in \numproduct{200x200} bins of the aforementioned parameter space.
The results of our detection-limit grid provide with several pieces of information. Firstly, with current data, we are already sensitive to the detection of additional Neptune-mass planets at periods up to \SI{e3}{\day}. Secondly, there
tend to be nearly no <10\% FAP signals for semi-amplitudes smaller than \SI{1}{\metre\per\second} (Fig.~\ref{fig:model_detectability}a). This is in line with what we expect from the uncertainty of LBL RV data. Thirdly, we observe a significantly suppressed detection near $P_\text{rot}$ and $P_\text{rot}/2$. This result, although expected, is very accentuated in our case: the detection limit goes to \SI{3}{\metre\per\second} near $P_\text{rot}$ and degrades to even higher values for $P_\text{rot}/2$. Finally, we observe what seems to be a forest of narrow drops in detection between \SIrange{1}{4}{\day}. This may be related to the particular WF of our dataset.

All discussed planets comfortably lie in the <10\% FAP zone of the parameter space -- and the posterior of GJ~3998~c is only narrowly missed by the detection drop associated with $P_\text{rot}/2$. Generally, detection drops can be mitigated in regimes where stellar-activity contributions in RV are not as strong. In this regard, there have been some indications that velocimetry in the near infrared may be a way in the right direction (e.g. \citealp{Carmona2023}). In any case, we must remark that our detection-limit test is pessimistic because it adheres to the assumptions of model d4 and the posteriors we obtained from its inference. 

\subsection{Stellar activity of GJ~3998}\label{sec:stellar_activity}
\citetalias{Affer2016} measured
\mbox{$\log R'_\text{HK}=5.01$} for GJ~3998, which implied \mbox{$P_\text{rot}=34.7\pm 6.9\,\unit\day$}
through the relation in \citet{SuarezMascareno2015}.
More recently, \citet{SuarezMascareno2018}, hereafter \citetalias{SuarezMascareno2018}, reported
\mbox{$P_\text{rot}=33.6\pm 3.6\,\unit\day$}. In our best model (d4; RV\&FWHM), we measure
\mbox{$P_\text{rot}=30.2\pm 0.3\,\unit\day$},
which agrees with \citetalias{SuarezMascareno2018}, but somewhat disagrees with \citetalias{Affer2016}
\mbox{($31.8^{+0.6}_{-0.5}\,\unit\day$)}.
The mismatch with \citetalias{Affer2016} is likely caused by our consideration of FWHM measurements together with RVs -- and ultimately, we work with more data than what was available at the time. In the variations of our best model for different activity indicators (Sect.~\ref{sec:rv_activity_indicators}; Table~\ref{tab:planet_recovery_indicators}), we measure similar values of $P_\text{rot}$:
\mbox{$30.2\pm 0.2\,\unit\day$} for RV\&S-index,
\mbox{$30.1\pm 0.3\,\unit\day$} for RV\&H$\alpha$, and
\mbox{$29.9\pm 0.3\,\unit\day$} for RV\&Na~I.

The timescales of active regions do not surprise either -- we report $\tau$ values of:
\mbox{$180^{+93}_{-55}\,\unit\day$} for the main RV\&FWHM model (d4),
\mbox{$321^{+278}_{-121}\,\unit\day$} for RV\&S-index,
\mbox{$130^{+52}_{-34}\,\unit\day$} for RV\&H$\alpha$, and
\mbox{$153^{+65}_{-41}\,\unit\day$} for RV\&Na~I. The ratio $\tau/P_\text{rot}$ appears to be roughly between 4 and 11, depending on the activity indicator. Therefore, GJ~3998 would likely have a light-curve morphology that is not Sun-like, as is defined in \citet{Giles2017}. The same work analysed the Kepler photometry of about 2200 stars with rotation periods near \SI{10}{\day} or \SI{20}{\day}. The authors found the following relation between active-region timescales and photometric rms:
\begin{align}
    \log_{10}\tau &= 10.9252 + 3.0123\log_{10}\text{rms} \nonumber \\
    &+0.5062(\log_{10}\text{rms})^2 -1.3606\log_{10}T_\text{eff},
    \label{eq:giles_relation}
\end{align}
where rms is measured in \unit\mag. Taking the confidence interval \mbox{$\tau=180^{+93}_{-55}\,\unit\day$} from model d4, using $T_\text{eff}$ from Table~\ref{tab:stellar_parameters} and solving for the rms gives an interval of \SIrange{11}{21}{\milli\mag}. Although \eqref{eq:giles_relation} was derived from shorter-period stars, this rms interval somewhat compares with the raw ASAS-SN photometry (\SI{15}{\milli\mag} rms in V, \SI{34}{\milli\mag} rms in g; Appendix~\ref{sec:photometry_analysis}).

The sinescales of active regions tend to be similar and span between \numrange{0.59}{0.93}. We observe positive $A_0,$ $B_0$, $A_1$ for all RV-indicator pairs. This suggests that stellar-activity modulations correlate positively across RV and activity indicators. Moreover, $A_0\ll B_0$ in all cases, which hints that stellar activity manifests itself more clearly in the RV gradient. This likely corresponds to a flux effect in what we are observing. The positive value of $B_0$ indicates that active regions are dark spots on the stellar surface.

Our data suggest no major differences between activity indicators in the short-term stellar activity of GJ~3998. The same could be stated for the long-term activity of GJ~3998, but with slightly reduced confidence. In our main model (d4), we measure what appears to be a sine component in the LTF, with a period \mbox{$P_\text{cyc}=316^{+14}_{-8}\,\unit\day$}. This result disagrees with \citetalias{SuarezMascareno2018} (\mbox{$1.8\pm 0.4\,\unit\year$}). To obtain this result, \citetalias{SuarezMascareno2018} inferred double-sinusoidal models at $P$ and $P/2$, following \citet{Berdyugina2005}. We note that the ratio between our and the \citetalias{SuarezMascareno2018} $P_\text{cyc}$ is \num{0.48(11)}. This could mean that their double-sinusoidal model had potentially picked up the strong \SI{316}{\day} signal, but had potentially assigned it to $P_\text{cyc}/2$ -- which would lead to the report of a twice as large period. This sine component in the LTF could correspond to a magnetic cycle that would be very short compared to other stars in the literature (e.g. \citealp{SuarezMascareno2016a,Distefano2017}; \citetalias{SuarezMascareno2018}). Other activity indicators inform us of a similar result. In our main model (d4), the sine component in the LTF has an RV amplitude of
$1.17^{+0.36}_{-0.37}\,\unit{\metre\per\second}$ and an FWHM amplitude of
$2.17^{+1.04}_{-1.12}\,\unit{\metre\per\second}$. We observe a somewhat significant amplitude in FWHM, but not so much in other indicators -- yet, our results suggest that the sine component in FWHM, H$\alpha$ and Na~I are in phase.\footnote{S-index phase posteriors also appear to be in phase, but should not be trusted since their underlying distribution is flat (Sect.~\ref{sec:rv_activity_indicators}).} This is compatible with the observed strong correlation between those indicators in the raw data (Fig.~\ref{fig:harpsn_correlations}).

Finally, we fitted one-dimensional variants of d4 on each activity indicator for completeness. Table~\ref{tab:1d_activity_posteriors} supplies with their posteriors. They are overall noisier than the ones in Table~\ref{tab:planet_recovery_indicators}, but we found no differences that merit further discussion.

\section{Conclusions}\label{sec:conclusions}
We measure GJ~3998~b, GJ~3998~c, and GJ~3998~d to have orbital periods of
\mbox{$2.65033^{+(22)}_{-(19)}$\,\unit\day},
\mbox{$13.727^{+0.003}_{-0.004}$\,\unit\day}, and
\mbox{$41.78\pm 0.05\,\unit\day$}, respectively;
minimum masses of
\mbox{$2.50^{+0.30}_{-0.29}$\,\unit\earthmass},
\mbox{$6.82^{+0.78}_{-0.75}$\,\unit\earthmass}, and
\mbox{$6.07^{+1.00}_{-0.96}$\,\unit\earthmass}, respectively;
and incident fluxes of
\mbox{$48^{+11}_{-9}\,\unit\earthflux$}, 
\mbox{$5.4^{+1.3}_{-1.0}\,\unit\earthflux$}, and
\mbox{$1.2^{+0.3}_{-0.2}\,\unit\earthflux$}, respectively. Our results paint a picture of a multi-planet system of super-Earths, where the outermost companion receives about as much incident flux as the Earth. This new discovery, GJ~3998~d, is one of the few currently known planets that come close to having an Earth-like incident flux. As such, it is a welcome addition to the statistics of planetary systems in the solar neighbourhood.

We did a chain of independent tests on the planetary nature of all three planets in the system. These tests include an alias analysis of activity peaks near orbital periods, as well as model comparisons between different: (i) stellar-activity kernels, (ii) planetary configurations, (iii) activity indicators, and (iv) RV extraction pipelines. In addition, we looked for planetary-like features in activity indicators, in case our stellar-activity modelling failed to account for such potential features. All planets passed our testing suite, and their posteriors demonstrated stability in different settings. Our detection-limit test shows that if a fourth planet existed, but somehow escaped detection, it would have likely induced an RV semi-amplitude smaller than \SI{1}{\metre\per\second} (Fig.~\ref{fig:model_detectability}). Current HARPS-N data suggests that GJ~3998 may exhibit magnetic activity with a period of
\mbox{$316^{+14}_{-8}$\,\unit\day}, which would place it among the stars with rather short magnetic-cycle periods. This longer-term stellar activity is well observed in three activity indicators: FWHM, H$\alpha$, and Na~I. Finally, we provide a tighter constraint on the rotational period of GJ~3998:
\mbox{$P_\text{rot}=30.2\pm 0.3\,\unit\day$}.

\begin{acknowledgements}
AKS acknowledges the support of a fellowship from the ``la Caixa'' Foundation (ID 100010434). The fellowship code is LCF/BQ/DI23/11990071.
AKS, JIGH, ASM, NN and RR acknowledge financial support from the Spanish Ministry of Science and Innovation (MICINN) project PID2020-117493GB-I00 and from the Government of the Canary Islands project ProID2020010129.
NN acknowledges funding from Light Bridges for the Doctoral Thesis ``Habitable Earth-like planets with ESPRESSO and NIRPS'',
in cooperation with the Instituto de Astrofísica de Canarias, and the use of Indefeasible Computer Rights (ICR) being commissioned at the ASTRO POC project in Tenerife, Canary Islands, Spain. The ICR-ASTRONOMY used for his research was provided by Light Bridges in cooperation with Hewlett Packard Enterprise (HPE).
LA and GM acknowledge support from the agreement ASI-INAF 2021-5-HH.2-2024.
IR and MP acknowledge fincancial support from Spanish grants PID2021-125627OB-C31 funded by MCIU/AEI/10.13039/501100011033 and by ``ERDF A way of making Europe'', PID2020-120375GB-I00 funded by MCIU/AEI, by the programme Unidad de Excelencia María de Maeztu CEX2020-001058-M, and by the Generalitat de Catalunya/CERCA programme.
IR acknowledges further financial support from the European Research Council (ERC) under the European Union’s Horizon Europe  programme (ERC Advanced Grant SPOTLESS; no. 101140786).
MP acknowledge support from the European Union -- NextGenerationEU (PRIN MUR 2022 20229R43BH) and the ``Programma di Ricerca Fondamentale INAF 2023''.
EGÁ acknowledges financial support from the Universidad Complutense de Madrid and the Spanish Ministerio de Ciencia e Innovación through the project PID2022-137241NB-C4[1,4].
JM and LA acknowledge support from the Italian Ministero dell'Università e della Ricerca and the European Union - Next Generation EU through project PRIN 2022 PM4JLH ``Know your little neighbours: characterising low-mass stars and planets in the Solar neighbourhood''.
AKS gratefully dedicates this work to V.~D.~Mihov for teaching him the principles of programming beyond mere coding.

This research has made use of NASA's Astrophysics Data System Bibliographic Services. We made use of NASA’s Astrophysics Data System, which is operated by the California Institute of Technology, under contract with the National Aeronautics and Space Administration under the Exoplanet Exploration Program.
This research has made use of the \textsc{SIMBAD} \citep{simbad} and \textsc{VizieR} \citep{vizier} databases, both operated at CDS, Strasbourg, France.
This work has made use of data from the European Space Agency (ESA) mission {\it Gaia} (\url{https://www.cosmos.esa.int/gaia}), processed by the {\it Gaia} Data Processing and Analysis Consortium (DPAC,
\url{https://www.cosmos.esa.int/web/gaia/dpac/consortium}). Funding for the DPAC has been provided by national institutions, in particular the institutions participating in the {\it Gaia} Multilateral Agreement.
We used the following \textsc{Python} packages for data analysis and
visualisation:
\textsc{Astropy} \citep{astropy},
\textsc{Matplotlib} \citep{matplotlib},
\textsc{nieva} (Stefanov et al., in prep.),
\textsc{NumPy} \citep{numpy},
\textsc{pandas} \citep{pandas1,pandas2},
\textsc{SciPy} \citep{scipy},
\textsc{seaborn} \citep{seaborn},
\textsc{s+leaf} \citep{spleaf1,spleaf2} and
\textsc{ultranest} \citep{ultranest}.
This manuscript was written and compiled in \textsc{Overleaf}. All presented analysis was conducted on \textsc{Ubuntu} machines. The bulk of modelling and inference was done on the Diva cluster (192 Xeon E7-4850 \SI{2.1}{\giga\hertz} CPUs; \SI{4.4}{\tera\byte} RAM) at Instituto de Astrofísica de Canarias, Tenerife, Spain.  \end{acknowledgements}

\bibliographystyle{aa} \bibliography{zotero_bibtex} 

\let\cleardoublepage\clearpage
\begin{appendix}
\section{Analysis of ASAS-SN photometry}\label{sec:photometry_analysis}
We used the ASAS-SN Sky Patrol online service\footnote{\url{https://asas-sn.osu.edu/}} to inspect the photometry of GJ~3998 at hand. At the time of access, the field of our target had been covered by ASAS-SN from October 2012 to June 2024 (\SIrange{2456200}{2460500}{\julianday}). Our preliminary check revealed that there are no significant contaminators in the field that could bias photometry. However, the total proper motion of GJ~3998 approaches \SI{0.4}{\arcsec\per\year} \citepalias{gaiaDR3}, which may become comparable to the pixel size of the detector (\SI{8}{\arcsec}). Past works addressed this issue through a computation of many separate light curves associated with smaller temporal intervals (e.g. \citealp{Trifonov2021,Damasso2023}). We went forwards with a similar procedure. We did split the region \SIrange{2456000}{2460500}{\julianday}, including endpoints, in timestamps at every \SI{300}{\julianday}. Then, for each time stamp, we computed the expected position of GJ~3998 and used the Sky Patrol service to compute an individual light curve for the temporal region defined by $\pm\SI{150}{\julianday}$ around the time stamp. Finally, we concatenated the light curves to a single data product, and performed iterative $3\sigma$ clipping by flux for each bandpass. Our final dataset contains 1110 measurements in Johnson V and 3335 photometric measurements in Sloan g.

\begin{figure}
    \centering
    \resizebox{\hsize}{!}{\includegraphics{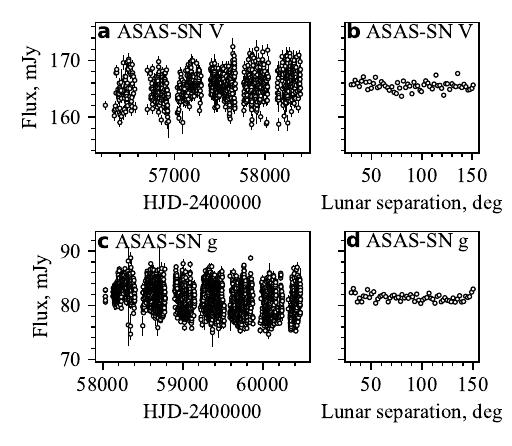}}
    \caption{
    (a,c) GJ~3998 ASAS-SN photometry in two bands: Johnson V and Sloan g. (b,d) The same time series against the lunar separation at the times of measurement, binned every \SI{2}{\degree}.
    }
    \label{fig:asassn_timeseries}
\end{figure}

Figure~\ref{fig:asassn_timeseries} displays the final combined ASAS-SN time series, as well as a binned scatter plot between flux and lunar separation to assess the level of lunar contamination. The rms of measurements against the median photometric uncertainty is \SI{2.3}{\milli\Jansky} against \SI{0.9}{\milli\Jansky} for Johnson V; and \SI{2.5}{\milli\Jansky} against \SI{0.5}{\milli\Jansky} for Sloan g. There are visible long-term trends in both bandpasses -- we observe a gradual increase in flux in Jonhson V and a likewise decrease in Sloan g. We calculated the minimum lunar separation throughout observations to be about \SI{29.6}{\deg}. Nevertheless, we do not observe appreciable differences in flux in low-separation regimes (Fig.~\ref{fig:asassn_timeseries}b,d). We thus attribute the long-term changes to an incomplete proper-motion correction.

\begin{figure*}
    \centering
    \resizebox{\hsize}{!}{\includegraphics{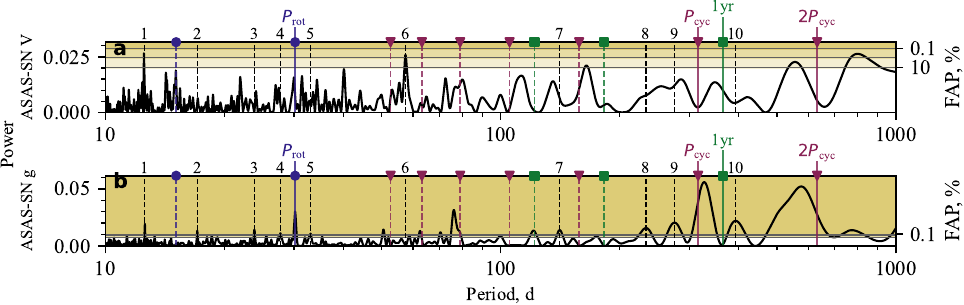}}
    \caption{
    GLSPs of GJ~3998 ASAS-SN photometry in: (a) Johnson V, (b) Sloan g. Both periodograms are saturated with many peaks. Some of them relate to the stellar rotation period $P_\text{rot}$ (solid blue line; circle), the sinusoidal cycle $P_\text{cyc}$ in the LTF (solid red line; downward triangle) to \SI{1}{\year} (solid green line; square). We plot the first few harmonics of each of these signals in dashed lines of a corresponding colour and marker. After this initial filtering, ten significant signals remain:
    (1) \SI{12.6}{\day}, unidentified;
    (2) \SI{17.1}{\day}, unidentified;
    (3) \SI{23.8}{\day}, unidentified;
    (4) \SI{27.7}{\day}, near the common harmonic of $P_\text{rot}$ and \SI{1}{\year};
    (5) \SI{33}{\day}, near the common harmonic of $P_7$ and \SI{1}{\year};
    (6) \SI{57}{\day}, unidentified;
    (7) \SI{141}{\day}, unidentified;
    (8) \SI{233}{\day}, near the common harmonic of $P_7$ and \SI{1}{\year};
    (9) \SI{276}{\day}, near $2P_7$;
    (10) \SI{393}{\day}, near $3P_7$.
    }
    \label{fig:asassn_gls}
\end{figure*}

Figure~\ref{fig:asassn_gls} gives the \SIrange{10}{1000}{\day} GLSPs of our time series in different bands. Before us stands a bouquet of $P_\text{rot}$, $P_\text{cyc}$ and \SI{1}{\year} harmonics, together with some higher-order harmonics of known and unknown periodicities (see labelled signals in Fig.~\ref{fig:asassn_gls}). Compared to the stellar activity we measure in RV (Sect.~\ref{sec:stellar_activity}), Johnson V photometry has insignificant $P_\text{rot}$ signals, a somewhat significant $2P_\text{cyc}$, and two strong signatures that we could not identify: at \SI{12.6}{\day} and at \SI{57}{\day}. On the other hand, Sloan g photometry contains strong $P_\text{rot}$ and $P_\text{cyc}$ signatures, as well as many unidentified peaks. Those include the \SI{12.6}{\day} and \SI{57}{\day} peaks from Johnson V, another \SI{23.8}{\day} peak, but most notably, a mysterious \SI{141}{\day} signal (hereafter $P_7$; see Fig.~\ref{fig:asassn_gls}). This last signal appears to comes to together with the signatures of $2P_7$ and $3P_7$; as well as its two yearly harmonics at \SI{33}{\day} and \SI{233}{\day}.

\begin{figure*}
    \centering
    \resizebox{\hsize}{!}{\includegraphics{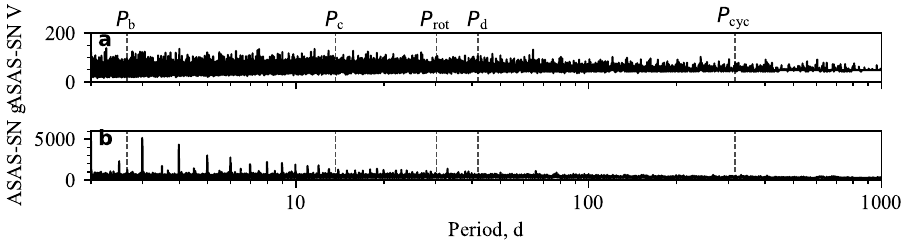}}
    \caption{
    BLSPs of GJ~3998 ASAS-SN photometry in: (a) Johnson V, (b) Sloan g.
    }
    \label{fig:asassn_bls}
\end{figure*}

What remains to be verified -- and what should not be affected by a long-term proper-motion drift -- is whether there are transit-like structures in the time series. We constructed box least-squares periodograms (BLSPs; \citealp{bls}) in the period interval between \SIrange{2}{1000}{\day} and a transit-duration interval between \SIrange{0.01}{1}{\day}. Figure~\ref{fig:asassn_bls} displays the result of this exercise. We find no significant peaks at the periodicities of any planet, nor at our measured rotational period $P_\text{rot}$ or long-term cycle $P_\text{cyc}$. We folded the time series for our best-model $P_\text{b}, P_\text{c}, P_\text{d}$ (Table~\ref{tab:bestmodel_posterior}), but we could not find no apparent structures. Repeating the exercise without the final step of sigma-clipping also gives a negative result. We conclude that current ASAS-SN data supports the argument of non-transiting GJ~3998~b, GJ~3998~c and GJ~3998~d.

\newpage
\section{Alias analysis of planets b and c}\label{sec:alias_analysis_bc}
\begin{figure*}
    \centering
    \resizebox{\hsize}{!}{\includegraphics{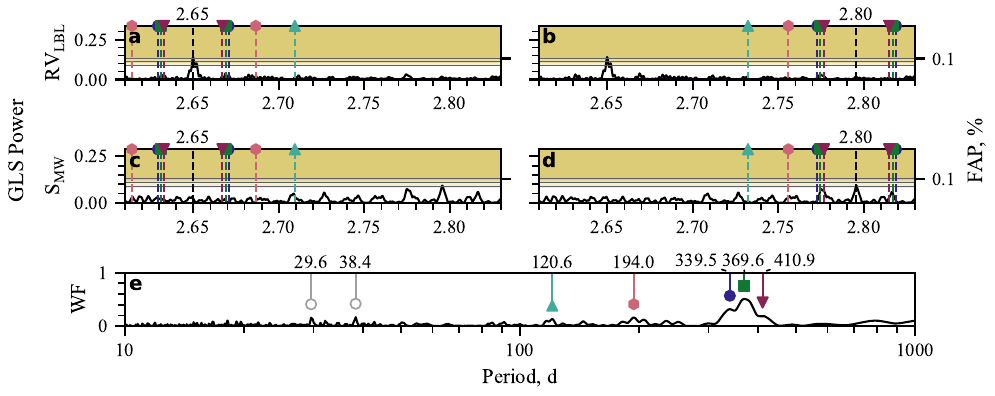}}
    \caption{
    Alias analysis of long-term detrended signals near \SI{2.65}{\day} in:
    (a,b)~LBL RV,
    (c,d)~S-index.
    The left column shows aliases of the \SI{2.65}{\day} signal, found in RV and attributed to GJ~3998~b. The right column shows aliases of a \SI{2.80}{\day} signal found in S-index.
    (e)~The WF reveals five prominent peaks at:
    \SI{120.6}{\day} (upward teal triangle),
    \SI{194.0}{\day} (pink hexagon),
    \SI{339.5}{\day} (blue circle),
    \SI{369.6}{\day} (green square) and
    \SI{410.9}{\day} (downward red triangle).
    Relevant aliases in data periodograms are plotted with dashed lines of a corresponding colour and marker.
    There are other strong peaks in the WF that do not create aliases in the specified period range (hollow grey circles).
    }
    \label{fig:alias_analysis_b}
\end{figure*}

We begin our alias analysis on the neighbourhood around GJ~3998~b. There are two peaks which we test for dependency: the planetary signal in RV (\SI{2.65}{\day}; Fig.~\ref{fig:harpsn_pgrams}c) and the \SI{2.80}{\day} signal in S-index (Fig.~\ref{fig:harpsn_pgrams}m). We select the most significant WF peaks in the interval \SIrange{10}{1000}{\day}. Three strong WF peaks come close to
\sfrac{1}{3}\,\unit{\year},
\sfrac{1}{2}\,\unit{\year} and
\SI{1}{\year}; these are
\SI{120.6}{\day},
\SI{194.0}{\day}, and
\SI{369.6}{\day},
respectively. For the two peaks of interest, they generate aliases within \SIrange{2.61}{2.83}{\day}. We select these intervals from the long-term detrended GLSPs and show them in Fig.~\ref{fig:alias_analysis_b}. Each signal comes with its \SI{1}{\year} aliases in its own dimension. The aliases of the \SI{2.80}{\day} signal are not even remotely close to the orbital period of GJ~3998~b. Equivalently, the \SI{2.65}{\day} planetary signal does not seem to form aliases near \SI{2.80}{\day}.

\begin{figure*}
    \centering
    \resizebox{\hsize}{!}{\includegraphics{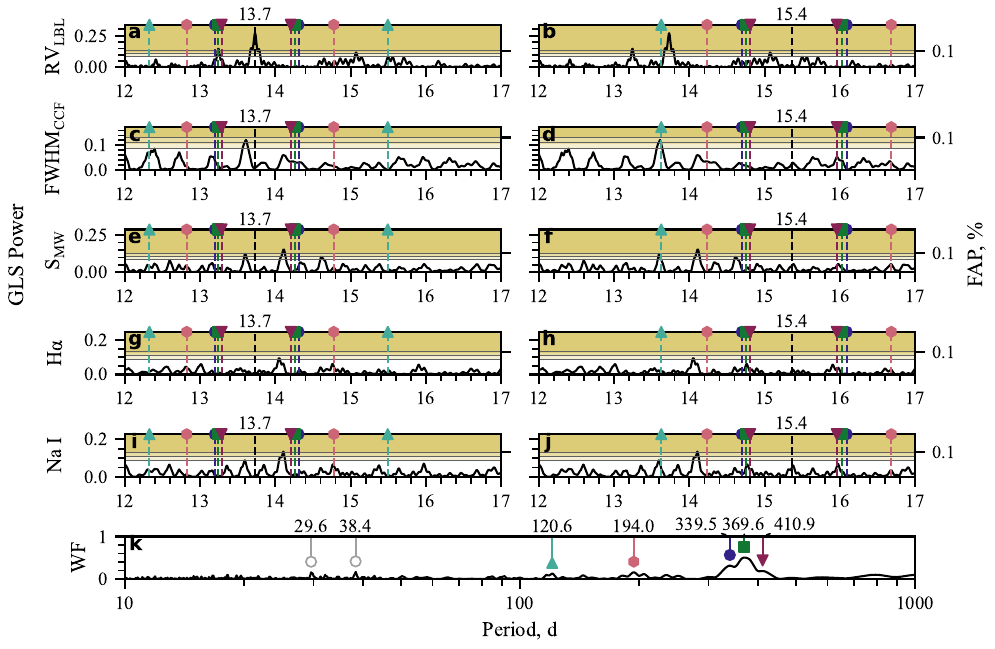}}
    \caption{
    Alias analysis of long-term detrended signals near \SI{13.7}{\day} in:
    (a,b)~LBL RV,
    (c,d)~CCF FWHM,
    (e,f)~S-index,
    (g,h)~H$\alpha$,
    (i,j)~Na~I.
    The left column shows aliases of the \SI{13.7}{\day} signal, found in RV and attributed to GJ~3998~c. The right column shows aliases of half the $P_\text{rot}$ peak found in the raw-data GLSP, at \SI{15.4}{\day}.
    (k)~The WF reveals five prominent peaks at:
    \SI{120.6}{\day} (upward teal triangle),
    \SI{194.0}{\day} (pink hexagon),
    \SI{339.5}{\day} (blue circle),
    \SI{369.6}{\day} (green square) and
    \SI{410.9}{\day} (downward red triangle).
    Relevant aliases in data periodograms are plotted with dashed lines of a corresponding colour and marker.
    There are other strong peaks in the WF that do not create aliases in the specified period range (hollow grey circles).
    }
    \label{fig:alias_analysis_c}
\end{figure*}

We continue with GJ~3998~c (\SI{13.7}{\day}) which is surrounded by strong signals in all activity indicators. We found out $P_\text{rot}/2$ has a \sfrac{1}{3}\,\unit{\year} alias near \SI{13.6}{\day}, close to the orbital period of GJ~3998~c. This is indeed what we observe in Fig.~\ref{fig:alias_analysis_c}: the aforementioned alias fits well peaks near the orbital period of the planet (Figs.~\ref{fig:alias_analysis_c}d,f,j; upward teal triangle) which do not agree exactly with the RV signal (compare Fig.~\ref{fig:alias_analysis_c}a with Figs.~\ref{fig:alias_analysis_c}c,e,i). There is another strong \SI{14.1}{\day} signal in S-index, which cannot be well explained by aliases of $P_\text{rot}/2$ (Fig.~\ref{fig:alias_analysis_c}f). This periodicity, however, is quite far away from GJ~3998~c (\SI{13.7}{\day}; Fig.~\ref{fig:alias_analysis_c}e).

  
\section{Supplementary material}

\begin{figure}
    \centering
    \resizebox{\hsize}{!}{\includegraphics{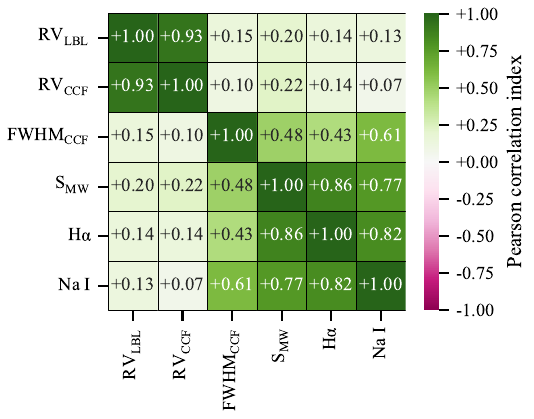}}
    \caption{
    Pearson correlation matrix between LBL RV, CCF RV, and discussed activity indicators.
    }
    \label{fig:harpsn_correlations}
\end{figure} \begin{figure}
    \centering
    \resizebox{\hsize}{!}{\includegraphics{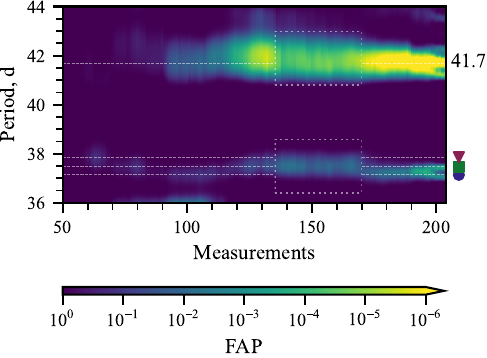}}
    \caption{
    FAP evolution of long-term detrended RV signals between \SIrange{36}{44}{\day}. We plot the location of the \SI{41.7}{\day} signal and its aliases within the interval. Alias symbols match
    Fig.~\ref{fig:alias_analysis_b}e,
    Fig.~\ref{fig:alias_analysis_c}k and
    Fig.~\ref{fig:alias_analysis_d}c.
    Dashed white rectangles display the power transfer from the \SI{41.7}{\day} signal to its \SIrange{37}{38}{\day} alias family.
    }
    \label{fig:harpsn_alias_fap_evolution}
\end{figure}

\begin{figure}
    \centering
    \resizebox{\hsize}{!}{\includegraphics{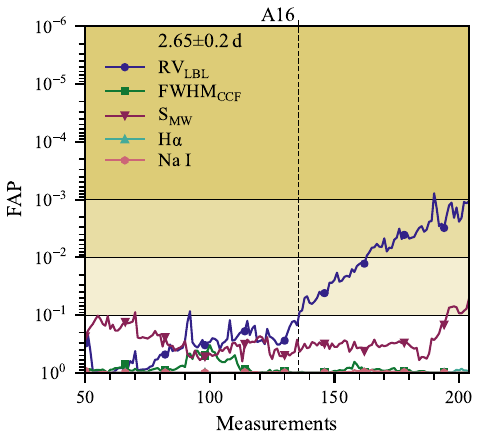}}
    \caption{
    FAP evolution of the strongest peak within \SI{0.2}{\day} of the \SI{2.65}{\day} signal for long-term detrended:
    LBL RV (blue circles),
    CCF FWHM (green squares),
    S-index (downward red triangles),
    H$\alpha$ (upward teal triangles) and
    Na~I (pink hexagons). The temporal coverage of \citetalias{Affer2016} is marked  by a dashed black line.
    }
    \label{fig:harpsn_fap_evolution_b}
\end{figure} \begin{figure}
    \centering
    \resizebox{\hsize}{!}{\includegraphics{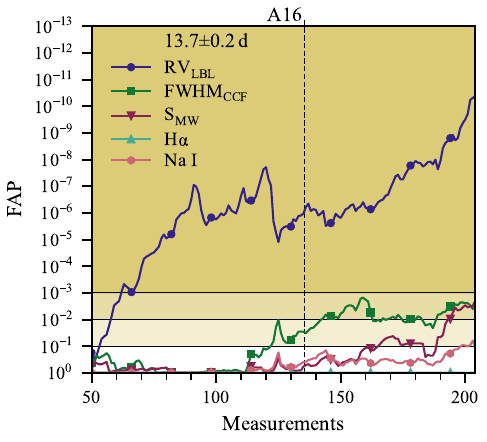}}
    \caption{
    FAP evolution of the strongest peak within \SI{0.2}{\day} of the \SI{13.7}{\day} signal for long-term detrended:
    LBL RV (blue circles),
    CCF FWHM (green squares),
    S-index (downward red triangles),
    H$\alpha$ (upward teal triangles) and
    Na~I (pink hexagons). The temporal coverage of \citetalias{Affer2016} is marked  by a dashed black line.
    }
    \label{fig:harpsn_fap_evolution_c}
\end{figure} 
\begin{figure*}
    \centering
    \resizebox{\hsize}{!}{\includegraphics{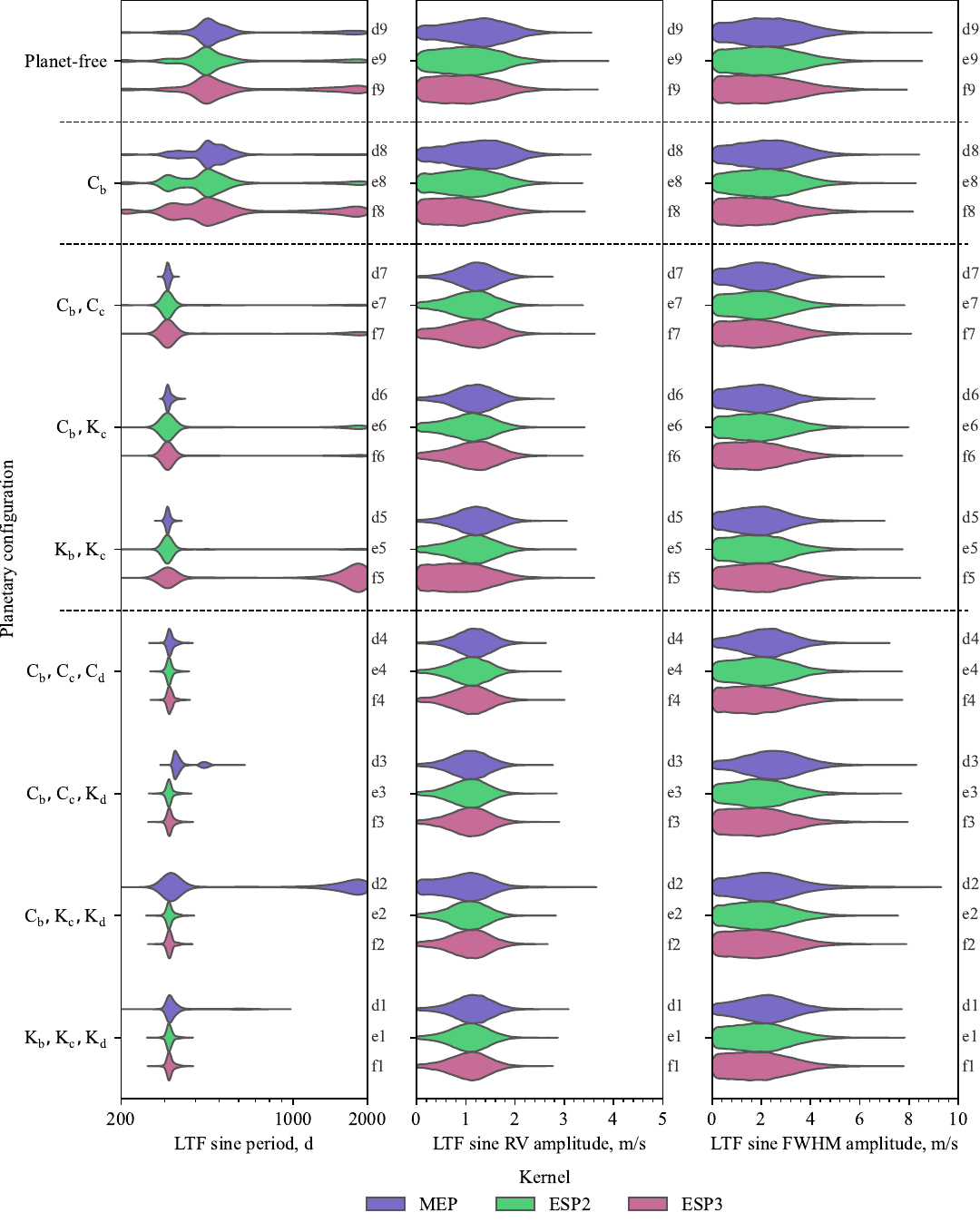}}
    \caption{
    Violin plots of all models with a sine component in their TFs that show the posteriors of:
    (a) the cycle period $P_\text{cyc}$,
    (b) the RV amplitude $k_\text{cyc, 0}$,
    (c) the RV amplitude $k_\text{cyc, 0}$.
    Violin plots are arranged vertically by planetary configuration in the same order as in Fig.~\ref{fig:model_table}. Plots come in three colours: blue for MEP, green for ESP2 and red for ESP3. Every distribution is accompanied by its corresponding model label.
    }
    \label{fig:model_corrector_period_distro}
\end{figure*}
\clearpage
\begin{table*}
\centering
\caption{
Period posteriors of GJ~3998~d in units of \unit{\day} from the grid-search in Fig.~\ref{fig:model_table}.
}
\label{tab:model_comparison_planetd_period}

\begin{tabular}{C{12pt}C{12pt}C{12pt}m{60pt}m{60pt}m{60pt}m{60pt}m{60pt}m{60pt}}
\hline\hline
\multicolumn{3}{c}{Planetary configuration} & \multicolumn{3}{c}{Sine-free LTF} & \multicolumn{3}{c}{Sine in LTF} \\ \cline{4-9}
b & c & d & \multicolumn{1}{c}{MEP} & \multicolumn{1}{c}{ESP2} & \multicolumn{1}{c}{ESP3} & \multicolumn{1}{c}{MEP} & \multicolumn{1}{c}{ESP2} & \multicolumn{1}{c}{ESP3} \\ \hline
C & C & C & $41.77\pm 0.05$&$41.77\pm 0.04$&$41.77\pm 0.04$&$41.78\pm 0.05$&$41.78^{+0.05}_{-0.04}$&$41.78^{+0.05}_{-0.04}$ \\
C & C & K & $41.78^{+0.07}_{-0.05}$&$41.77^{+0.05}_{-0.04}$&$41.77^{+0.05}_{-0.04}$&$41.78^{+0.06}_{-0.05}$&$41.77^{+0.05}_{-0.04}$&$41.78^{+0.05}_{-0.04}$ \\
C & K & K & $41.78^{+0.08}_{-0.05}$&$41.77^{+0.05}_{-0.04}$&$41.77^{+0.05}_{-0.04}$&$41.79^{+0.09}_{-0.06}$&$41.77^{+0.05}_{-0.04}$&$41.77^{+0.05}_{-0.04}$ \\
K & K & K & $41.78^{+0.08}_{-0.05}$&$41.77^{+0.05}_{-0.04}$&$41.77^{+0.05}_{-0.04}$&$41.78^{+0.06}_{-0.05}$&$41.77^{+0.05}_{-0.04}$&$41.78^{+0.05}_{-0.04}$ \\ \hline
\end{tabular}
\tablefoot{
Reported uncertainties reflect the 16\textsuperscript{th} and the 84\textsuperscript{th} percentiles.
}
\end{table*}

\begin{table*}
\centering
\caption{
Phase posteriors of GJ~3998~d from the grid-search in Fig.~\ref{fig:model_table}.
}
\label{tab:model_comparison_planetd_phase}
\begin{tabular}{C{12pt}C{12pt}C{12pt}m{60pt}m{60pt}m{60pt}m{60pt}m{60pt}m{60pt}}
\hline\hline
\multicolumn{3}{c}{Planetary configuration} & \multicolumn{3}{c}{Sine-free LTF} & \multicolumn{3}{c}{Sine in LTF} \\ \cline{4-9}
b & c & d & \multicolumn{1}{c}{MEP} & \multicolumn{1}{c}{ESP2} & \multicolumn{1}{c}{ESP3} & \multicolumn{1}{c}{MEP} & \multicolumn{1}{c}{ESP2} & \multicolumn{1}{c}{ESP3} \\ \hline
C & C & C & $0.79\pm 0.08$&$0.79\pm 0.07$&$0.79\pm 0.07$&$0.79\pm 0.08$&$0.79\pm 0.07$&$0.79\pm 0.07$ \\
C & C & K & $0.75^{+0.09}_{-0.12}$&$0.77^{+0.08}_{-0.09}$&$0.77\pm 0.08$&$0.76^{+0.09}_{-0.11}$&$0.77\pm 0.08$&$0.77^{+0.07}_{-0.08}$ \\
C & K & K & $0.75^{+0.09}_{-0.14}$&$0.77\pm 0.08$&$0.77\pm 0.08$&$0.74^{+0.10}_{-0.19}$&$0.77\pm 0.08$&$0.77\pm 0.08$ \\
K & K & K & $0.75^{+0.09}_{-0.14}$&$0.77^{+0.08}_{-0.09}$&$0.77^{+0.08}_{-0.09}$&$0.76^{+0.09}_{-0.11}$&$0.77\pm 0.08$&$0.77\pm 0.08$ \\ \hline
\end{tabular}
\tablefoot{
Reported uncertainties reflect the 16\textsuperscript{th} and the 84\textsuperscript{th} percentiles.
}
\end{table*}

\begin{table*}
\centering
\caption{
Radial-velocity semi-amplitude posteriors of GJ~3998~d in units of \unit{\metre\per\second} from the grid-search in Fig.~\ref{fig:model_table}.
}
\label{tab:model_comparison_planetd_krv}
\begin{tabular}{C{12pt}C{12pt}C{12pt}m{60pt}m{60pt}m{60pt}m{60pt}m{60pt}m{60pt}}
\hline\hline
\multicolumn{3}{c}{Planetary configuration} & \multicolumn{3}{c}{Sine-free LTF} & \multicolumn{3}{c}{Sine in LTF} \\ \cline{4-9}
b & c & d & \multicolumn{1}{c}{MEP} & \multicolumn{1}{c}{ESP2} & \multicolumn{1}{c}{ESP3} & \multicolumn{1}{c}{MEP} & \multicolumn{1}{c}{ESP2} & \multicolumn{1}{c}{ESP3} \\ \hline
C & C & C & $1.76^{+0.27}_{-0.28}$&$1.84\pm 0.24$&$1.85\pm 0.24$&$1.74^{+0.26}_{-0.25}$&$1.79\pm 0.23$&$1.79^{+0.24}_{-0.25}$ \\
C & C & K & $1.81^{+0.30}_{-0.29}$&$1.91\pm 0.26$&$1.90\pm 0.26$&$1.81^{+0.28}_{-0.27}$&$1.83\pm 0.24$&$1.82\pm 0.25$ \\
C & K & K & $1.81^{+0.31}_{-0.29}$&$1.90^{+0.26}_{-0.25}$&$1.90^{+0.27}_{-0.26}$&$1.79^{+0.32}_{-0.30}$&$1.84^{+0.25}_{-0.24}$&$1.82\pm 0.25$ \\
K & K & K & $1.82^{+0.32}_{-0.30}$&$1.91^{+0.26}_{-0.25}$&$1.91\pm 0.26$&$1.78\pm 0.28$&$1.83\pm 0.24$&$1.82\pm 0.25$ \\ \hline
\end{tabular}
\tablefoot{
Reported uncertainties reflect the 16\textsuperscript{th} and the 84\textsuperscript{th} percentiles.
}
\end{table*}

\begin{table*}
\centering
\caption{
Eccentricity posteriors of GJ~3998~d from the grid-search in Fig.~\ref{fig:model_table}.
}
\label{tab:model_comparison_planetd_e}
\begin{tabular}{C{12pt}C{12pt}C{12pt}m{60pt}m{60pt}m{60pt}m{60pt}m{60pt}m{60pt}}
\hline\hline
\multicolumn{3}{c}{Planetary configuration} & \multicolumn{3}{c}{Sine-free LTF} & \multicolumn{3}{c}{Sine in LTF} \\ \cline{4-9}
b & c & d & \multicolumn{1}{c}{MEP} & \multicolumn{1}{c}{ESP2} & \multicolumn{1}{c}{ESP3} & \multicolumn{1}{c}{MEP} & \multicolumn{1}{c}{ESP2} & \multicolumn{1}{c}{ESP3} \\ \hline
C & C & C & - & - & - & - & - & - \\
C & C & K & $0.16^{+0.22}_{-0.12}$&$0.18^{+0.16}_{-0.12}$&$0.17^{+0.16}_{-0.11}$&$0.14^{+0.19}_{-0.10}$&$0.13^{+0.14}_{-0.09}$&$0.12^{+0.14}_{-0.09}$ \\
C & K & K & $0.17^{+0.23}_{-0.12}$&$0.18^{+0.16}_{-0.12}$&$0.17^{+0.16}_{-0.12}$&$0.18^{+0.30}_{-0.13}$&$0.13^{+0.15}_{-0.09}$&$0.12^{+0.14}_{-0.09}$ \\
K & K & K & $0.17^{+0.24}_{-0.12}$&$0.18^{+0.16}_{-0.12}$&$0.17^{+0.16}_{-0.12}$&$0.13^{+0.19}_{-0.10}$&$0.13^{+0.14}_{-0.09}$&$0.13^{+0.14}_{-0.09}$ \\ \hline
\end{tabular}
\tablefoot{
Reported uncertainties reflect the 16\textsuperscript{th} and the 84\textsuperscript{th} percentiles. Three-circular models did not sample for this parameter.
}
\end{table*}

\begin{table*}
\centering
\caption{
Argument-of-periastron posteriors of GJ~3998~d in units of \unit{\rad} from the grid-search in Fig.~\ref{fig:model_table}.
}
\label{tab:model_comparison_planetd_w}
\begin{tabular}{C{12pt}C{12pt}C{12pt}m{60pt}m{60pt}m{60pt}m{60pt}m{60pt}m{60pt}}
\hline\hline
\multicolumn{3}{c}{Planetary configuration} & \multicolumn{3}{c}{Sine-free LTF} & \multicolumn{3}{c}{Sine in LTF} \\ \cline{4-9}
b & c & d & \multicolumn{1}{c}{MEP} & \multicolumn{1}{c}{ESP2} & \multicolumn{1}{c}{ESP3} & \multicolumn{1}{c}{MEP} & \multicolumn{1}{c}{ESP2} & \multicolumn{1}{c}{ESP3} \\ \hline
C & C & C & - & - & - & - & - & - \\
C & C & K & $2.36^{+1.78}_{-0.91}$&$2.33^{+0.98}_{-0.75}$&$2.33^{+1.12}_{-0.83}$&$2.49^{+2.09}_{-1.02}$&$2.58^{+1.68}_{-1.00}$&$2.61^{+1.69}_{-1.09}$ \\
C & K & K & $2.36^{+1.73}_{-0.88}$&$2.31^{+0.98}_{-0.74}$&$2.32^{+1.11}_{-0.82}$&$2.38^{+1.65}_{-0.86}$&$2.53^{+1.58}_{-1.00}$&$2.59^{+1.74}_{-1.06}$ \\
K & K & K & $2.33^{+1.65}_{-0.86}$&$2.31^{+1.00}_{-0.77}$&$2.31^{+1.14}_{-0.79}$&$2.52^{+1.96}_{-1.15}$&$2.61^{+1.72}_{-1.05}$&$2.55^{+1.68}_{-1.06}$ \\ \hline
\end{tabular}
\tablefoot{
Reported uncertainties reflect the 16\textsuperscript{th} and the 84\textsuperscript{th} percentiles. Three-circular models did not sample for this parameter.
}
\end{table*} \clearpage

\begin{table*}[h]
\centering
\caption{
Parameter posteriors of our best model (d4), against the results published by \citetalias{Affer2016}.
}
\label{tab:bestmodel_posterior}

\begin{tabular}{lcclcc}
\hline \hline
Parameter name &
Symbol &
Unit &
Prior &
\multicolumn{2}{c}{Posterior} \\ \cline{5-6}
 &
 &
 &
 &
This work &
\citetalias{Affer2016} \\ \hline
\textbf{Long-term function parameters} &
 &
 &
 &
 &
 \\
Period &
$P_\text{cyc}$ &
\unit\day &
$\mathcal{U}_{\log}\left(200, 2000\right)$ &
$316^{+14}_{-8}$ &
- \\
RV phase &
$\varphi_\text{cyc, 0}$ &
- &
$\mathcal{U}\left(0,1\right)$\tablefootmark{w} &
$0.571^{+0.255}_{-0.257}$ &
- \\
RV amplitude &
$k_\text{cyc, 0}$ &
\unit{\metre\per\second} &
$\mathcal{U}\left(0,40\right)$ &
$1.17^{+0.36}_{-0.37}$ &
- \\
FWHM phase &
$\varphi_\text{cyc, 1}$ &
- &
$\mathcal{U}\left(0,1\right)$\tablefootmark{w} &
$0.812^{+0.247}_{-0.259}$ &
- \\
FWHM amplitude &
$k_\text{cyc, 1}$ &
\unit{\metre\per\second} &
$\mathcal{U}\left(0,40\right)$ &
$2.17^{+1.04}_{-1.12}$ &
- \\
RV second-order correction &
$\alpha_0$ &
\unit{\metre\per\second\per\day\squared} &
$\mathcal{N}(\mu_\text{lm},200\sigma_\text{lm})$ &
$(-3.25^{+4.37}_{-4.56})\times\num{e-7}$ &
- \\
RV first-order correction &
$\beta_0$ &
\unit{\metre\per\second\per\day} &
$\mathcal{N}(\mu_\text{lm},200\sigma_\text{lm})$ &
$(-6.71^{+173.14}_{-179.03})\times\num{e-5}$ &
- \\
RV zero-order correction &
$\gamma_0$ &
\unit{\metre\per\second} &
$\mathcal{N}(\mu_\text{lm},200\sigma_\text{lm})$ &
$2.48^{+1.69}_{-1.70}$ &
$-0.27^{+0.49}_{-0.43}$ \\
FWHM second-order correction &
$\alpha_1$ &
\unit{\metre\per\second\per\day\squared} &
$\mathcal{N}(\mu_\text{lm},200\sigma_\text{lm})$ &
$(1.37^{+1.59}_{-1.63})\times\num{e-6}$ &
- \\
FWHM first-order correction &
$\beta_1$ &
\unit{\metre\per\second\per\day} &
$\mathcal{N}(\mu_\text{lm},200\sigma_\text{lm})$ &
$(5.45^{+6.15}_{-6.38})\times\num{e-3}$ &
- \\
FWHM zero-order correction &
$\gamma_1$ &
\unit{\metre\per\second} &
$\mathcal{N}(\mu_\text{lm},200\sigma_\text{lm})$ &
$3.27^{+5.66}_{-5.61}$ &
- \\
\textbf{Dataset parameters} &
 &
 &
 &
 &
\\
RV jitter &
$J_{0,0}$ &
\unit{\metre\per\second} &
$\mathcal{U}_{\log}\left(10^{-3}, 10^{3}\right)$ &
$1.29^{+0.15}_{-0.14}$ &
$1.19^{+0.11}_{-0.14}$ \\
FWHM jitter &
$J_{0,1}$ &
\unit{\metre\per\second} &
$\mathcal{U}_{\log}\left(10^{-3}, 10^{3}\right)$ &
$0.03^{+0.29}_{-0.03}$ &
- \\
\textbf{Stellar-activity parameters} &
 &
 &
 &
 &
 \\
Timescale &
$\tau$ &
\unit\day &
$\mathcal{U}_{\log}\left(40, 10^{4}\right)$ &
$180^{+93}_{-55}$ &
$34.4^{+11.6}_{-2.0}$ \\
Period &
$P_\text{rot}$ &
\unit\day &
$\mathcal{U_{\log}}\left(2, 200\right)$ &
$30.2\pm 0.3$ &
$31.8^{+0.6}_{-0.5}$ \\
Sinescale (harmonic complexity) &
$\eta$ &
- &
$\mathcal{U}_{\log}\left(10^{-2}, 10^{2}\right)$ &
$0.83^{+0.29}_{-0.19}$ &
$0.41^{+0.05}_{-0.04}$ \\
RV amplitude &
$A_0$ &
\unit{\metre\per\second} &
$\mathcal{U}(-10^3,10^3)$ &
$1.32^{+0.70}_{-0.44}$ &
$3.07^{+0.29}_{-0.24}$ \\
RV gradient amplitude &
$B_0$ &
\unit{\metre\per\second\per\day} &
$\mathcal{U}(-10^3,10^3)$ &
$18.1^{+8.2}_{-4.4}$ &
- \\
FWHM amplitude &
$A_1$ &
\unit{\metre\per\second} &
$\mathcal{U}_{\log}\left(10^{-3}, 10^{3}\right)$ &
$6.03^{+2.54}_{-1.41}$ &
- \\
\textbf{GJ~3998~b} &
 &
 &
 &
 &
 \\
Period &
$P_\text{b}$ &
\unit\day &
$\mathcal{U}\left(2, 3\right)$ &
$2.65033^{+(22)}_{-(19)}$ &
$2.64977^{+(81)}_{-(77)}$ \\
Phase &
$\varphi_\text{b}$ &
- &
$\mathcal{U}\left(0, 1\right)$\tablefootmark{w} &
$0.148^{+0.082}_{-0.094}$ &
- \\
Inferior-conjunction ephemeris &
$\varepsilon_\text{b}$ &
\unit\julianday$-\num{2400000}$ &
derived &
$60202.964^{+0.249}_{-0.216}$ &
$56905.895^{+0.042}_{-0.040}$ \\
RV semi-amplitude &
$k_\text{rv, b}$ &
\unit{\metre\per\second} &
$\mathcal{U}\left(0, 5\right)$ &
$1.79\pm 0.18$ &
$1.82^{+0.14}_{-0.16}$ \\
Eccentricity &
$e_\text{b}$ &
- &
- &
0 &
0 \\
Semi-major axis &
$a_\text{b}$ &
\unit\au &
derived &
$0.030\pm 0.001$ &
$0.029\pm 0.001$ \\
Minimum mass &
$m_\text{b}\sin i_\text{b}$ &
\unit\earthmass &
derived &
$2.50^{+0.30}_{-0.29}$ &
$2.47\pm 0.27$  \\
Incident flux &
$\Phi_\text{b}$ &
\unit\earthflux &
derived &
$48^{+11}_{-9}$ &
- \\
\textbf{GJ~3998~c} &
 &
 &
 &
 &
 \\
Period &
$P_\text{c}$ &
\unit\day &
$\mathcal{U}\left(13, 14\right)$ &
$13.727^{+0.003}_{-0.004}$ &
$13.740\pm 0.016$ \\
Phase &
$\varphi_\text{c}$ &
- &
$\mathcal{U}\left(0, 1\right)$\tablefootmark{w} &
$0.466^{+0.055}_{-0.051}$ &
- \\
Inferior-conjunction ephemeris &
$\varepsilon_\text{c}$ &
\unit\julianday$-\num{2400000}$ &
derived &
$60196.96^{+0.70}_{-0.75}$ &
$56902.2^{+0.24}_{-0.29}$ \\
RV semi-amplitude &
$k_\text{rv, c}$ &
\unit{\metre\per\second} &
$\mathcal{U}\left(0, 5\right)$ &
$2.86\pm 0.26$ &
$2.67^{+0.28}_{-0.23}$ \\
Eccentricity &
$e_\text{c}$ &
- &
- &
0 &
$0.049^{+0.052}_{-0.034}$ \\
Semi-major axis &
$a_\text{c}$ &
\unit\au &
derived &
$0.090\pm 0.003$ &
$0.089\pm 0.003$ \\
Minimum mass &
$m_\text{c}\sin i_\text{c}$ &
\unit\earthmass &
derived &
$6.82^{+0.78}_{-0.75}$ &
$6.26^{+0.79}_{-0.76}$ \\
Incident flux &
$\Phi_\text{c}$ &
\unit\earthflux &
derived &
$5.4^{+1.3}_{-1.0}$ &
- \\
\textbf{GJ~3998~d} &
 &
 &
 &
 &
 \\
Period &
$P_\text{d}$ &
\unit\day &
$\mathcal{U}\left(30, 50\right)$ &
$41.78\pm 0.05$ &
- \\
Phase &
$\varphi_\text{d}$ &
- &
$\mathcal{U}\left(0, 1\right)$\tablefootmark{w} &
$0.785^{+0.076}_{-0.080}$ &
-  \\
Inferior-conjunction ephemeris &
$\varepsilon_\text{d}$ &
\unit\julianday$-\num{2400000}$ &
derived &
$60170.55^{+3.29}_{-3.15}$ &
- \\
RV semi-amplitude &
$k_\text{rv, d}$ &
\unit{\metre\per\second} &
$\mathcal{U}\left(0, 5\right)$ &
$1.74^{+0.26}_{-0.25}$ &
- \\
Eccentricity &
$e_\text{d}$ &
- &
- &
0 &
- \\
Semi-major axis &
$a_\text{d}$ &
\unit\au &
derived &
$0.189\pm 0.006$ &
- \\
Minimum mass &
$m_\text{d}\sin i_\text{d}$ &
\unit\earthmass &
derived &
$6.07^{+1.00}_{-0.96}$ &
- \\
Incident flux &
$\Phi_\text{d}$ &
\unit\earthflux &
derived &
$1.2^{+0.3}_{-0.2}$ &
- \\[1ex] \hline
\end{tabular}
\tablefoot{
\tablefoottext{w}{Wrapped parameter.}
Reported uncertainties reflect the 16\textsuperscript{th} and the 84\textsuperscript{th} percentiles. $P_\text{b}$ uncertainties are given in parentheses, to the order of the least significant figure. Some LTF parameters have priors informed by an initial Levenberg-Marquadt fitting procedure.
}
\end{table*} \clearpage

\begin{figure*}
    \centering
    \resizebox{\hsize}{!}{\includegraphics{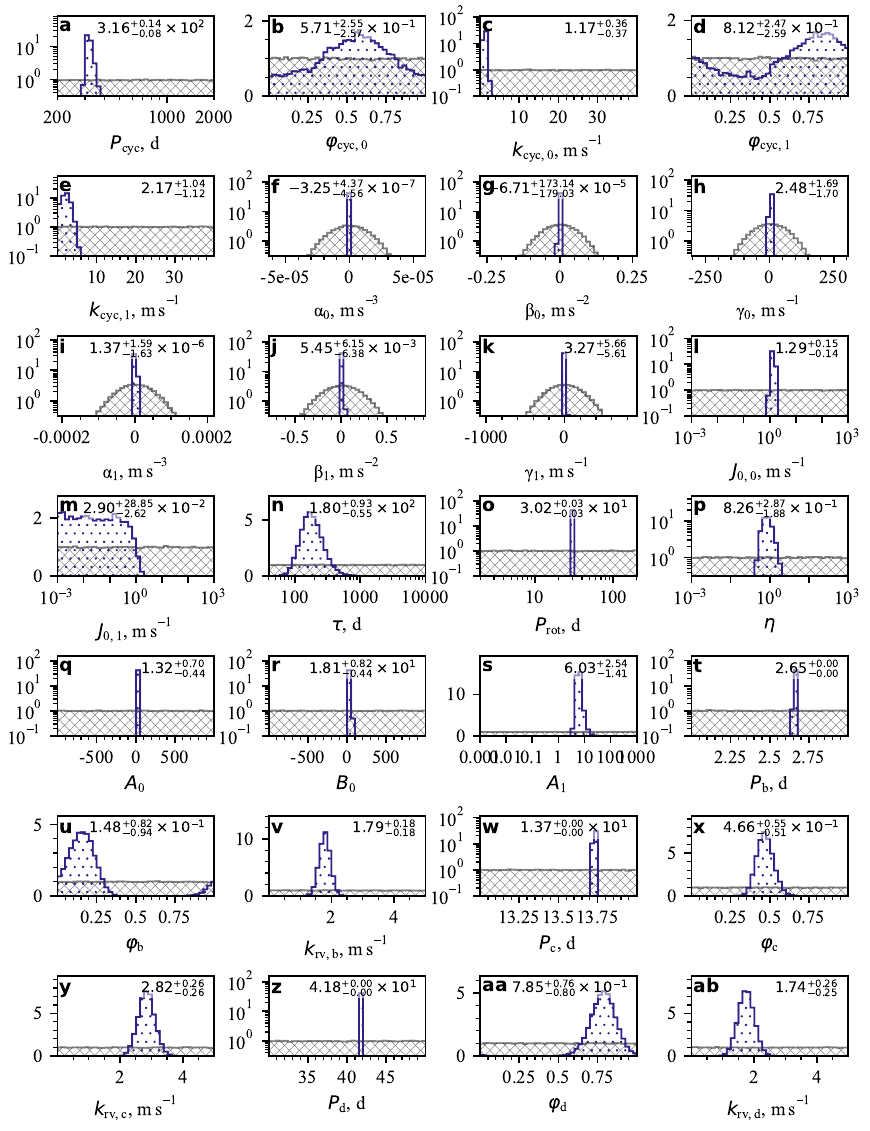}}
    \caption{Visual extension of Table~\ref{tab:bestmodel_posterior}: prior (grey, cross-hatched) and posterior (blue, dot-hatched) distributions of all parameters in our best model (d4). Uncertainties in our posteriors reflect the 16\textsuperscript{th} and the 84\textsuperscript{th} percentiles. For the sake of brevity, we show three significant digits for posterior medians, and truncate uncertainties to match the same precision.
    }
    \label{fig:bestmodel_posteriors}
\end{figure*} 
\begin{sidewaystable*}[h]
\small
\centering
\caption{
Comparison of posteriors from modelling of different RVs and activity indicators. These are:
CCF RV \& FWHM;,
LBL RV \& CCF FWHM (d4, our best model; boldface),
LBL RV \& S-index,
LBL RV \& H$\alpha$,
LBL RV \& Na I.
}
\label{tab:planet_recovery_indicators}

\begin{tabular}{lcclccccc}
\hline \hline
\multirow{2}{*}{Parameter name} &
\multirow{2}{*}{Symbol} &
\multirow{2}{*}{Unit} &
\multicolumn{1}{c}{\multirow{2}{*}{Prior}} &
\multicolumn{5}{c}{Posterior} \\ \cline{5-9} 
 &
 &
 &
 &
RV\textsubscript{CCF} \& FWHM\textsubscript{CCF} &
\textbf{RV\textsubscript{LBL} \& FWHM\textsubscript{CCF}} &
RV\textsubscript{LBL} \& S-index &
RV\textsubscript{LBL} \& H$\alpha$ &
RV\textsubscript{LBL} \& Na I \\ \hline
\textbf{Long-term function parameters} &
 &
 &
 &
 &
 &
 &
 \\
Period &
$P_\text{cyc}$ &
\unit\day &
$\mathcal{U}_{\log}\left(200, 2000\right)$ &
$1844^{+110}_{-162}$ &
$316^{+14}_{-8}$ &
$332^{+47}_{-19}$ &
$283^{+104}_{-13}$ &
$1384^{+502}_{-945}$ \\
RV phase &
$\varphi_\text{cyc, 0}$ &
- &
$\mathcal{U}\left(0,1\right)$\tablefootmark{w} &
$0.507^{+0.194}_{-0.156}$ &
$0.643^{+0.350}_{-0.230}$ &
$0.690^{+0.403}_{-0.302}$ &
$0.736^{+0.276}_{-0.233}$ &
$0.723^{+0.264}_{-0.236}$ \\
RV amplitude &
$k_\text{cyc, 0}$ &
\unit{\metre\per\second} &
$\mathcal{U}\left(0,40\right)$ &
$1.77^{+1.19}_{-1.10}$ &
$1.17^{+0.36}_{-0.37}$ &
$1.07^{+0.36}_{-0.41}$ &
$1.27^{+0.30}_{-0.31}$ &
$0.95^{+0.55}_{-0.61}$ \\
Indicator phase &
$\varphi_\text{cyc, 1}$ &
- &
$\mathcal{U}\left(0,1\right)$\tablefootmark{w} &
$0.932^{+0.194}_{-0.197}$ &
$0.882^{+0.333}_{-0.226}$ &
$0.00^{+0.37}_{-0.35}$ &
$0.984^{+0.243}_{-0.257}$ &
$0.994^{+0.244}_{-0.224}$ \\
Indicator amplitude &
$k_\text{cyc, 1}$ &
- &
$\mathcal{U}_{\log}\left(10^{-3}, 10^{3}\right)$ &
$3.42^{+2.78}_{-2.28}$ &
$2.17^{+1.04}_{-1.12}$ &
($3.76^{+2.26}_{-2.20})\times 10^{-2}$ &
($5.51^{+5.68}_{-3.82})\times 10^{-3}$ &
($2.07^{+1.30}_{-1.36})\times 10^{-3}$ \\
RV second-order correction &
$\alpha_0$ &
\unit{\metre\per\second\per\day\squared} &
$\mathcal{N}(\mu_\text{lm},200\sigma_\text{lm})$ &
($1.48^{+11.03}_{-9.75})\times 10^{-7}$ &
($-3.25^{+4.37}_{-4.56})\times 10^{-7}$ &
($-6.74^{+6.27}_{-7.25})\times 10^{-7}$ &
($-6.63^{+3.71}_{-3.76})\times 10^{-7}$ &
($-5.34^{+6.24}_{-3.96})\times 10^{-7}$ \\
RV first-order correction &
$\beta_0$ &
\unit{\metre\per\second\per\day} &
$\mathcal{N}(\mu_\text{lm},200\sigma_\text{lm})$ &
($1.35^{+4.03}_{-3.75})\times 10^{-3}$ &
($-6.71^{+173.14}_{-179.03})\times 10^{-5}$ &
($-1.47^{+2.42}_{-2.79})\times 10^{-3}$ &
($-1.29^{+1.42}_{-1.43})\times 10^{-3}$ &
($-6.25^{+20.67}_{-14.99})\times 10^{-4}$ \\
RV zero-order correction &
$\gamma_0$ &
\unit{\metre\per\second} &
$\mathcal{N}(\mu_\text{lm},200\sigma_\text{lm})$ &
$3.00^{+3.30}_{-3.29}$ &
$2.48^{+1.69}_{-1.70}$ &
$1.39^{+2.21}_{-2.47}$ &
$2.04^{+1.34}_{-1.50}$ &
$2.58^{+1.40}_{-1.53}$ \\
Indicator second-order correction &
$\alpha_1$ &
- &
$\mathcal{N}(\mu_\text{lm},200\sigma_\text{lm})$ &
($1.52^{+2.53}_{-2.16})\times 10^{-6}$ &
($1.37^{+1.59}_{-1.63})\times 10^{-6}$ &
($4.82^{+3.92}_{-4.26})\times 10^{-8}$ &
($7.05^{+4.67}_{-4.56})\times 10^{-9}$ &
($1.18^{+0.95}_{-1.00})\times 10^{-9}$ \\
Indicator first-order correction &
$\beta_1$ &
- &
$\mathcal{N}(\mu_\text{lm},200\sigma_\text{lm})$ &
($5.98^{+9.51}_{-8.60})\times 10^{-3}$ &
($5.45^{+6.15}_{-6.38})\times 10^{-3}$ &
($2.92^{+1.51}_{-1.66})\times 10^{-4}$ &
($4.84^{+1.81}_{-1.80})\times 10^{-5}$ &
($7.94^{+3.64}_{-3.89})\times 10^{-6}$ \\
Indicator zero-order correction &
$\gamma_1$ &
- &
$\mathcal{N}(\mu_\text{lm},200\sigma_\text{lm})$ &
$2.87^{+7.77}_{-7.29}$ &
$3.27^{+5.66}_{-5.61}$ &
($3.98^{+1.34}_{-1.42})\times 10^{-1}$ &
($7.54^{+1.63}_{-1.69})\times 10^{-2}$ &
($1.17^{+0.33}_{-0.34})\times 10^{-2}$ \\
\textbf{Dataset parameters} &
 &
 &
 &
 &
 &
 &
 \\
RV jitter &
$J_{0,0}$ &
\unit{\metre\per\second} &
$\mathcal{U}_{\log}\left(10^{-3}, 10^{3}\right)$ &
$1.46\pm 0.24$ &
$1.29^{+0.15}_{-0.14}$ &
$1.28^{+0.15}_{-0.14}$ &
$1.31^{+0.16}_{-0.15}$ &
$1.30^{+0.15}_{-0.14}$ \\
Indicator jitter &
$J_{0,1}$ &
- &
$\mathcal{U}_{\log}\left(10^{-3}, 10^{3}\right)$ &
($4.46^{+34.56}_{-4.12})\times 10^{-2}$ &
($2.90^{+28.85}_{-2.62})\times 10^{-2}$ &
($3.05^{+4.46}_{-1.65})\times 10^{-3}$ &
($9.59^{+0.70}_{-0.66})\times 10^{-3}$ &
($1.51^{+0.21}_{-0.21})\times 10^{-3}$ \\
\textbf{Stellar-activity parameters} &
 &
 &
 &
 &
 &
 &
 \\
Timescale &
$\tau$ &
\unit\day &
$\mathcal{U}_{\log}\left(40, 10^{4}\right)$ &
$180^{+119}_{-60}$ &
$180^{+93}_{-55}$ &
$321^{+278}_{-121}$ &
$130^{+52}_{-34}$ &
$153^{+65}_{-41}$ \\
Period &
$P$ &
\unit\day &
$\mathcal{U}_{\log}\left(2, 200\right)$ &
$30.2^{+0.4}_{-0.3}$ &
$30.2\pm 0.3$ &
$30.2\pm 0.2$ &
$30.1\pm 0.3$ &
$29.9\pm 0.3$ \\
Sinescale (harmonic complexity) &
$\eta$ &
- &
$\mathcal{U}_{\log}\left(10^{-2}, 10^{2}\right)$ &
$1.00^{+0.53}_{-0.28}$ &
$0.82^{+0.29}_{-0.19}$ &
$0.93^{+0.45}_{-0.25}$ &
$0.65^{+0.18}_{-0.13}$ &
$0.59^{+0.19}_{-0.14}$ \\
RV amplitude &
$A_0$ &
\unit{\metre\per\second} &
$\mathcal{U}(-10^3,10^3)$ &
$2.77^{+2.02}_{-0.88}$ &
$1.32^{+0.70}_{-0.44}$ &
$2.00^{+1.87}_{-0.77}$ &
$0.76^{+0.46}_{-0.47}$ &
$1.17^{+0.63}_{-0.68}$ \\
RV gradient amplitude &
$B_0$ &
\unit{\metre\per\second\per\day} &
$\mathcal{U}(-10^3,10^3)$ &
($2.56^{+1.90}_{-0.86})\times 10^{1}$ &
($1.81^{+0.82}_{-0.44})\times 10^{1}$ &
($2.30^{+1.89}_{-0.75})\times 10^{1}$ &
($1.53^{+0.46}_{-0.29})\times 10^{1}$ &
($1.45^{+0.49}_{-0.30})\times 10^{1}$ \\
Indicator amplitude &
$A_1$ &
- &
$\mathcal{U}_\text{log}(10^{-3},10^3)$ &
$7.09^{+4.88}_{-2.02}$ &
$6.03^{+2.54140}_{-1.40612}$ &
($1.38^{+1.08}_{-0.44})\times 10^{-1}$ &
($1.98^{+0.53}_{-0.35})\times 10^{-2}$ &
($3.65^{+1.02}_{-0.67})\times 10^{-3}$ \\
\textbf{GJ~3998~b} &
 &
 &
 &
 &
 &
 &
 \\
Period &
$P_\text{b}$ &
\unit\day &
$\mathcal{U}\left(2, 3\right)$ &
$2.65060^{+(120)}_{-(53)}$ &
$2.65033^{+(22)}_{-(19)}$ &
$2.65030^{+(22)}_{-(19)}$ &
$2.65029^{+(18)}_{-(16)}$ &
$2.65027^{+(17)}_{-(16)}$ \\
Phase &
$\varphi_\text{b}$ &
- &
$\mathcal{U}\left(0, 1\right)$\tablefootmark{w} &
$0.263^{+0.253}_{-0.236}$ &
$0.148^{+0.082}_{-0.095}$ &
$0.159^{+0.080}_{-0.098}$ &
$0.166^{+0.068}_{-0.076}$ &
$0.172^{+0.068}_{-0.075}$ \\
RV semi-amplitude &
$k_\text{rv, b}$ &
\unit{\metre\per\second} &
$\mathcal{U}\left(0, 5\right)$ &
$1.62\pm 0.24$ &
$1.79\pm 0.18$ &
$1.82\pm 0.18$ &
$1.84\pm 0.18$ &
$1.84\pm 0.18$ \\
\textbf{GJ~3998~c} &
 &
 &
 &
 &
 &
 &
 \\
Period &
$P_\text{c}$ &
\unit\day &
$\mathcal{U}\left(13, 14\right)$ &
$13.724^{+0.004}_{-0.005}$ &
$13.727^{+0.003}_{-0.004}$ &
$13.726\pm 0.004$ &
$13.727\pm 0.004$ &
$13.727\pm 0.004$ \\
Phase &
$\varphi_\text{c}$ &
- &
$\mathcal{U}\left(0, 1\right)$\tablefootmark{w} &
$0.511^{+0.073}_{-0.069}$ &
$0.466^{+0.055}_{-0.051}$ &
$0.483^{+0.058}_{-0.056}$ &
$0.492^{+0.058}_{-0.057}$ &
$0.495^{+0.062}_{-0.060}$ \\
RV semi-amplitude &
$k_\text{rv, c}$ &
\unit{\metre\per\second} &
$\mathcal{U}\left(0, 5\right)$ &
$2.92^{+0.34}_{-0.33}$ &
$2.82\pm 0.26$ &
$2.68\pm 0.25$ &
$3.01^{+0.31}_{-0.30}$ &
$2.73\pm 0.29$ \\
\textbf{GJ~3998~d} &
 &
 &
 &
 &
 &
 &
 \\
Period &
$P_\text{d}$ &
\unit\day &
$\mathcal{U}\left(30, 50\right)$ &
$41.77^{+0.11}_{-10.53}$ &
$41.78\pm 0.05$ &
$41.78\pm 0.05$ &
$41.78^{+0.06}_{-0.05}$ &
$41.79^{+0.06}_{-0.05}$ \\
Phase &
$\varphi_\text{d}$ &
- &
$\mathcal{U}\left(0, 1\right)$\tablefootmark{w} &
$0.753^{+0.221}_{-0.181}$ &
$0.785^{+0.077}_{-0.080}$ &
$0.781^{+0.086}_{-0.088}$ &
$0.785^{+0.078}_{-0.083}$ &
$0.770^{+0.084}_{-0.087}$ \\
RV semi-amplitude &
$k_\text{rv, d}$ &
\unit{\metre\per\second} &
$\mathcal{U}\left(0, 5\right)$ &
$1.52^{+0.41}_{-0.45}$ &
$1.74^{+0.26}_{-0.25}$ &
$1.69^{+0.26}_{-0.27}$ &
$1.90\pm 0.27$ &
$1.67^{+0.27}_{-0.28}$ \\
\hline
\end{tabular}
\tablefoot{
\tablefoottext{w}{Wrapped parameter.}
Reported uncertainties reflect the 16\textsuperscript{th} and the 84\textsuperscript{th} percentiles. $P_\text{b}$ uncertainties are given in parentheses, to the order of the least significant figure. Some LTF parameters have priors informed by an initial Levenberg-Marquadt fitting procedure.
Models with H$\alpha$ and Na~I used the prior $\mathcal{U}_{\log}\left(10^{-4}, 1\right)$ for their LTF amplitudes.
}
\end{sidewaystable*} 
\begin{sidewaystable*}[h]
\centering
\caption{
Comparison of posteriors from one-dimensional modelling of different activity indicators: CCF FWHM,
S-index,
H$\alpha$ and
Na~I.
}
\label{tab:1d_activity_posteriors}

\begin{tabular}{lcclcccc}
\hline \hline
\multirow{2}{*}{Parameter name} &
\multirow{2}{*}{Symbol} &
\multirow{2}{*}{Unit} &
\multicolumn{1}{c}{\multirow{2}{*}{Prior}} &
\multicolumn{4}{c}{Posterior} \\ \cline{5-8} 
 &
 &
 &
 &
FWHM\textsubscript{CCF} &
S-index &
H$\alpha$ &
Na I \\ \hline
\textbf{Long-term function parameters} &
 &
 &
 &
 &
 &
 &
 \\
Period &
$P_\text{cyc}$ &
\unit\day &
$\mathcal{U}_{\log}\left(200, 2000\right)$ &
$584^{+932}_{-264}$ &
$497^{+186}_{-30}$ &
$563^{+825}_{-188}$ &
$509^{+728}_{-71}$ \\
Indicator phase &
$\varphi_\text{cyc, 0}$ &
- &
$\mathcal{U}\left(0,1\right)$\tablefootmark{w} &
$0.887^{+0.316}_{-0.342}$ &
$0.286^{+0.304}_{-0.285}$ &
$0.023^{+0.267}_{-0.288}$ &
$0.997^{+0.223}_{-0.233}$ \\
Indicator amplitude &
$k_\text{cyc, 0}$ &
- &
$\mathcal{U}_{\log}\left(10^{-3}, 10^{3}\right)$ &
$1.86^{+1.69}_{-1.29}$ &
$(5.88^{+1.76}_{-2.54})\times 10^{-2}$ &
$(7.07^{+6.69}_{-5.00})\times 10^{-2}$ &
$(2.23^{+1.14}_{-1.45})\times 10^{-3}$ \\
Indicator second-order correction &
$\alpha_0$ &
- &
$\mathcal{N}(\mu_\text{lm},200\sigma_\text{lm})$ &
$(1.36^{+1.30}_{-1.37})\times 10^{-6}$ &
$(4.52^{+3.38}_{-3.55})\times 10^{-8}$ &
$(8.62^{+4.28}_{-4.48})\times 10^{-9}$ &
$(1.56^{+0.89}_{-0.93})\times 10^{-9}$ \\
Indicator first-order correction &
$\beta_0$ &
- &
$\mathcal{N}(\mu_\text{lm},200\sigma_\text{lm})$ &
$(5.43^{+5.07}_{-5.21})\times 10^{-3}$ &
$(2.85^{+1.42}_{-1.40})\times 10^{-4}$ &
$(5.47^{+1.71}_{-1.72})\times 10^{-5}$ &
$(9.50^{+3.43}_{-3.69})\times 10^{-6}$ \\
Indicator zero-order correction &
$\gamma_0$ &
- &
$\mathcal{N}(\mu_\text{lm},200\sigma_\text{lm})$ &
$3.06^{+4.85}_{-4.68}$ &
$(4.14^{+1.44}_{-1.30})\times 10^{-1}$ &
$(7.93^{+1.56}_{-1.68})\times 10^{-2}$ &
$(1.26\pm 0.32)\times 10^{-2}$ \\
\textbf{Dataset parameters} &
 &
 &
 &
 &
 &
 &
 \\
Indicator jitter &
$J_{0,0}$ &
- &
$\mathcal{U}_{\log}\left(10^{-3}, 10^{3}\right)$ &
$(2.69^{+24.04}_{-2.41})\times 10^{-2}$ &
$(2.88^{+4.00}_{-1.50})\times 10^{-3}$ &
$(9.27^{+0.68}_{-0.61})\times 10^{-3}$ &
$(1.37\pm 0.21)\times 10^{-3}$ \\
\textbf{Stellar-activity parameters} &
 &
 &
 &
 &
 &
 &
 \\
Timescale &
$\tau$ &
\unit\day &
$\mathcal{U}_{\log}\left(40, 10^{4}\right)$ &
$138^{+75}_{-44}$ &
$527^{+1507}_{-288}$ &
$116^{+55}_{-34}$ &
$128^{+59}_{-38}$ \\
Period &
$P$ &
\unit\day &
$\mathcal{U}_{\log}\left(2, 200\right)$ &
$30.9^{+0.8}_{-0.9}$ &
$29.6^{+0.3}_{-0.2}$ &
$29.6\pm 0.6$ &
$29.3^{+0.6}_{-0.7}$ \\
Sinescale (harmonic complexity) &
$\eta$ &
- &
$\mathcal{U}_{\log}\left(10^{-2}, 10^{2}\right)$ &
$0.71^{+0.35}_{-0.19}$ &
$0.78^{+0.85}_{-0.30}$ &
$0.70^{+0.25}_{-0.16}$ &
$0.61^{+0.25}_{-0.16}$  \\
Indicator amplitude &
$A_0$ &
- &
$\mathcal{U}_\text{log}(10^{-3},10^3)$ &
$5.10^{+1.96}_{-1.06}$ &
$(1.19^{+2.02}_{-0.44})\times 10^{-1}$ &
$(1.85^{+0.58}_{-0.33})\times 10^{-2}$ &
$(3.56^{+1.17}_{-0.72})\times 10^{-3}$  \\
\hline
\end{tabular}
\tablefoot{
\tablefoottext{w}{Wrapped parameter.}
Reported uncertainties reflect the 16\textsuperscript{th} and the 84\textsuperscript{th} percentiles. Some LTF parameters have priors informed by an initial Levenberg-Marquadt fitting procedure.
Models with H$\alpha$ and Na~I used the prior $\mathcal{U}_{\log}\left(10^{-4}, 1\right)$ for their LTF amplitudes.
}
\end{sidewaystable*}
\end{appendix}
\end{document}